\documentclass[aps,prd,twocolumn,reprint,preprintnumbers,superscriptaddress, nofootinbib]{revtex4-1}

\usepackage{natbib}
\usepackage{verbatim}
\usepackage{slashed} %To use Feynman slash notation
\usepackage{hyperref}
\usepackage{color}
\usepackage{amsmath}
\usepackage{amsfonts}
\usepackage{amssymb}
\usepackage[utf8]{inputenc}
\usepackage{graphicx}
\usepackage{microtype}

\def\nn{\nonumber}
\def\({\left(}
\def\){\right)}
\def\[{\left[}
\def\]{\right]}
\newcommand{\nc}{\newcommand}
\nc{\beq}{\begin{equation}}
\nc{\eeq}{\end{equation}}
\nc{\mc}{\mathcal}
\nc{\ddt}[1]{\frac{\partial #1}{\partial t}}
\nc{\ddx}[1]{\frac{\partial #1}{\partial \chi}}
\nc{\ddxx}[1]{\frac{\partial^2 #1}{\partial \chi^2}}
\nc{\rhoeff}{\rho_{\mathrm{eff}}}
\newcommand{\ud}{\mathrm{d}}

\definecolor{darkgreen}{cmyk}{0.85,0.2,1.00,0.2}

\usepackage{xcolor}

\begin{document}

\title{Stochastic evolution of scalar fields with continuous symmetries during inflation}

\author{Peter Adshead}
\email{adshead@illinois.edu}
\affiliation{Department of Physics, University of Illinois at Urbana-Champaign, Urbana, IL 61801, USA }

\author{Lauren Pearce}
\email{lpearce@psu.edu}
\affiliation{Pennsylvania State University-New Kensington, New Kensington, PA 15068 }

\author{Jessie Shelton}
\email{sheltonj@illinois.edu}
\affiliation{Department of Physics, University of Illinois at Urbana-Champaign, Urbana, IL 61801, USA }

\author{Zachary J. Weiner}
\email{zweiner2@illinois.edu}
\affiliation{Department of Physics, University of Illinois at Urbana-Champaign, Urbana, IL 61801, USA }

\begin{abstract}
During inflation, scalar fields with masses less than the Hubble scale acquire vacuum expectation values (vevs) via stochastic processes driven by quantum fluctuations. For nearly massless spectator scalars transforming nontrivially under a continuous symmetry group, we demonstrate that the evolution of the vev depends on the dimensionality of the scalar field space. Fields in larger representations both attain larger vacuum expectation values and converge more rapidly to equilibrium. We present an argument demonstrating how this higher-dimensional evolution can be obtained in unitary gauge for fields transforming under local symmetries with a mass gap that is small compared to the Hubble scale.  Finally, we show that accounting for the full number of degrees of freedom in the Standard Model Higgs multiplet tightens Higgs stability constraints on the inflationary scale at the percent level and has more dramatic consequences for both the vev and the energy stored in the Higgs field after inflation.
\end{abstract}

\maketitle

\section{Introduction}

The primordial power spectrum of density fluctuations as inferred from the Cosmic Microwave Background radiation (CMB) is broadly consistent with minimal single-field inflation~\cite{Akrami:2018odb}.  During inflation, quantum fluctuations of the inflaton and metric are stretched to superhorizon scales where they freeze-out to generate adiabatic, red-tilted spectra of curvature and gravitational wave fluctuations~\cite{Mukhanov:1981xt, Guth:1982ec, Hawking:1982cz, Bardeen:1983qw}.  Furthermore, any minimally-coupled scalar field with a mass that is small compared to the Hubble scale during inflation generically acquires a nonzero vacuum expectation value (vev); the generation of this vev and its evolution can be understood as a stochastic process driven by quantum fluctuations~\cite{Starobinsky:1986fx,Rey:1986zk,Sasaki:1987gy,Nambu:1988je,Morikawa:1989xz, Linde:1991sk, Starobinsky:1994bd}. In particular, absent direct couplings to the inflaton or curvature scalar, the Standard Model (SM) Higgs acquires a vev during inflation~\cite{DeSimone:2012qr, Kunimitsu:2012xx, Choi:2012cp, Enqvist:2013kaa, Enqvist:2014bua}. This breaking of electroweak symmetry, and its subsequent restoration during reheating, has been suggested as a means to produce the Standard Model~\cite{Figueroa:2015rqa,Figueroa:2016dsc,Figueroa:2017slm}, as a source for gravitational waves~\cite{Figueroa:2014aya,Figueroa:2016ojl,Espinosa:2018eve} and primordial black holes~\cite{Espinosa:2017sgp} (however, see Ref.~\cite{Passaglia:2019ueo}), and as a baryogenesis mechanism~\cite{Kusenko:2014lra,Pearce:2015nga,Yang:2015ida, Adshead:2015jza}.

One striking feature of the observed SM Higgs is the instability of its potential at high energy scales~\cite{Sher:1988mj,Sher:1993mf,Casas:1994qy,Altarelli:1994rb}. Given the observed Higgs and top quark masses, the running quartic coupling becomes negative at scales $\sim 10^{9} - 10^{13} \, \mathrm{GeV}$ including 2$\sigma$ uncertainties on $m_h, m_t$, and $\alpha_s(m_Z)$~\cite{Casas:1996aq,Isidori:2001bm,Isidori:2007vm,Ellis:2009tp,EliasMiro:2011aa, Degrassi:2012ry,Buttazzo:2013uya,Branchina:2013jra,Bednyakov:2015sca}. While the electroweak vacuum itself is metastable,
regions of space in which excursions of the Higgs vev venture sufficiently far into the unstable part of the potential during inflation can be fatal to our Universe.
The exclusion of such regions from our past lightcone was used to derive constraints on the allowable Hubble scale during inflation in Refs.~\cite{Espinosa:2015qea,Kearney:2015vba,East:2016anr,Kohri:2017iyl}.

Beyond the SM Higgs field (and possibly the inflaton), no other scalars are known to exist in nature. However, the existence of light scalars during the inflationary epoch has been routinely invoked for both fundamental and phenomenological reasons. On the fundamental side, scalar moduli are a common ingredient in realistic models of string cosmology (see, e.g., Ref.~\cite{Kane:2015jia} for a review), which typically yield a large spectrum of moduli, some of which may be light during inflation. On the phenomenological side, the curvaton mechanism relies on the quantum fluctuations from an additional light spectator field to subsequently generate the curvature perturbation following inflation~\cite{Linde:1996gt,Enqvist:2001zp,Lyth:2001nq,Moroi:2001ct, Lyth:2002my}.  The stochastic inflationary population of SM-singlet scalars has also been invoked as a population mechanism for dark matter~\cite{Chung:2001cb, Chung:2004nh,  Nurmi:2015ema, Enqvist:2017kzh, Alonso-Alvarez:2018tus, Markkanen:2018gcw, Tenkanen:2019aij}.  A similar mechanism may be used to produce cold axion dark matter~\cite{Preskill:1982cy,Abbott:1982af,Dine:1982ah,Graham:2018jyp,Guth:2018hsa}.

The stochastic evolution of scalar fields during inflation has primarily been studied for real, singlet scalars, following the pioneering treatment of Starobinsky and Yokoyama~\cite{Starobinsky:1994bd}. In this work, we extend the work of Starobinsky and Yokoyama to light scalar fields transforming linearly under a continuous symmetry group.  We demonstrate that the vev of  a scalar field that transforms under a continuous symmetry undergoes stochastic evolution in a field space whose dimensionality is determined by the number of real scalar degrees of freedom.  This higher-dimensional diffusion leads to larger asymptotic vevs, as well as faster evolution to the asymptotic probability distribution governing the vev.  We begin by considering global symmetries and subsequently discuss the situation when the symmetry is gauged.  As a useful example we consider a field moving in a quartic potential, which admits analytic solutions;  however, the techniques may be generalized, and we also present numerical results for the SM Higgs.

In describing the evolution of the Higgs vev, studies to date have implicitly assumed that only the radial mode is important during the random walk. That is, the Higgs has been treated as a single real scalar field. In this approach, the field range of the radial mode is often extended to unphysical negative values in order to maintain a $Z_2$ symmetry under which the origin of field space is stationary.  However, the SM Higgs is an $\mathrm{SU}(2)$ doublet with four degrees of freedom. Although the Higgs is subject to gauge transformations that can be chosen to make it appear one-dimensional away from the origin (i.e., unitary gauge), we argue that its field space is effectively four-dimensional during at least part of its random walk during inflation; see \cite{Hertzberg:2018kyi,Jain:2019wxo} for related arguments in a different context.  We show that accounting for the larger field space does not significantly impact the probability for the Higgs to attain catastrophically large vevs. However, the larger field space leads to order unity corrections to the value of the root-mean-square Higgs vev and the corresponding energy density in the Higgs field.

This paper is organized as follows. In Sec.~\ref{sec:vev}, we begin by developing the Fokker-Planck equation describing the stochastic evolution of a scalar field with a continuous global symmetry, beginning with $\mathrm{SO}(N)$ and then generalizing to other symmetry groups.  We solve the resulting Fokker-Planck equation analytically and numerically for a quartic potential in order to demonstrate three important consequences of the enlarged field space. In Sec.~\ref{sec:gauging}, we then discuss gauging the symmetry, and  in Sec.~\ref{sec:inflation}, we consider the implications for the stochastic evolution of the SM Higgs during inflation. We conclude in Sec.~\ref{sec:conclusion}, and provide details of our computations in appendices.  Appendix~\ref{sec:lneq0} considers the behavior of excited modes in the quartic potential, while Appendix~\ref{sec:comp} provides details of our numerical scheme for solving the Fokker-Planck equation.

We work in natural units where $\hbar = c = k_B = M_{\rm Pl} = 1$, and our convention for the metric is the `mostly minus' signature.

%%%%%%%%%%%%%%%%%%%%%%%
\section{Vacuum Expectation Value}
\label{sec:vev}
%%%%%%%%%%%%%%%%%%%%%%%

We consider a generic inflationary theory described by the action $S = S_{\rm EH}+S_{\rm inf}+S_{\rm spec}$, where gravity is described by the usual Einstein-Hilbert action $S_{\rm EH}$,  $S_{\rm inf}$ describes a sector that sources the (quasi-)de Sitter inflationary background, and $S_{\rm spec}$ describes a minimally coupled spectator scalar,
\begin{align}
S_{\rm spec} = \int \ud^4x \, \sqrt{-g} \left[ \frac{1}{2} (\partial \vec{\chi}) \cdot  (\partial \vec{\chi}) -V \left(  \vec{\chi} \cdot \vec{\chi} \right) \right].
\end{align}
We take the scalar $\vec{\chi}$ to transform under a continuous symmetry group that is respected by the potential.
Any couplings between $\vec{\chi}$ and the inflaton must typically be small to avoid spoiling slow-roll inflation, and in what follows we take them to be zero.

During inflation, $\vec\chi$ acquires a vev, $\langle \vec \chi\cdot \vec \chi\rangle\neq 0$.
The evolution of this vev can be described by a Fokker-Planck equation, where the stochastic noise driving the diffusion of the vev is provided by quantum fluctuations of subhorizon modes~\cite{Starobinsky:1994bd}.  We begin by generalizing the treatment of Ref.~\cite{Starobinsky:1994bd} to the scenario where $\vec{\chi}$ transforms in the fundamental representation of a global $\mathrm{SO}(N)$ symmetry before making the further generalization to scalar fields transforming under other continuous symmetry groups.

%%%%%%%%%%%%%%
\subsection{Global $\mathrm{SO}(N)$ Symmetry}
%%%%%%%%%%%%%%

Following the stochastic approach developed in Ref.~\cite{Starobinsky:1994bd}, we divide each component of the field $\vec{\chi}$ into long- and short-wavelength modes,
\begin{align}
\chi_i(t,\vec{x})& = \chi_{L,i}(t,\vec{x}) + \int \frac{\ud^3k}{(2\pi)^{3 / 2}}
\Theta (k-\epsilon a H) \nonumber \\
&\quad \phantom{spacerrrr} \times \left( a_{\vec{k},i} \chi_{\vec{k}} e^{-i \vec{k} \cdot \vec{x}}
+ a_{\vec{k},i}^\dagger \chi_{\vec{k}}^\dagger e^{i \vec{k} \cdot \vec{x}}\right),
\label{eq:long_short_division}
\end{align}
with the division occurring at $k = \epsilon a H$.  The long-wavelength modes, denoted by $\vec{\chi}_L$, can to good approximation be treated classically,
while the short-wavelength modes are quantized.  For the long-wavelength modes to be outside the horizon, we must take $\epsilon \lesssim 1$.\footnote{ To respect the $\mathrm{SO}(N)$ symmetry, the parameter $\epsilon$, which implements the division between long and short wavelength modes, must be identical for each component.}  For sufficiently small values of $\epsilon$, the short-wavelength modes with $k \gtrsim \epsilon a H $ satisfy the massless Klein-Gordon equation in de Sitter space, with solution
\begin{align}
\label{eq:wavefns}
\chi_{\vec{k}} = \frac{H}{\sqrt{2k}} \left( \tau - \frac{i}{k} \right) e^{-ik\tau},
\end{align}
where $\tau = -1/aH$ is conformal time, and $H = \dot{a}/a$ is the Hubble rate. Here and throughout, overdots represent derivatives with respect to cosmic time $t$.

The equation of motion for the long-wavelength modes,
\begin{align}
\label{eq:Langevin}
\dot{\chi}_{L,i}(t,\vec{x}) = - \frac{1}{3H} \frac{d V}{d \chi_{L,i}} + f_i(t,\vec{x}),
\end{align}
can be viewed as a Langevin equation with the stochastic source term $f_i$ generated by the ``freezing out'' of short-wavelength modes,
\begin{align}
f_i &= \int \frac{\ud^3k}{(2\pi)^{3 / 2}}
\delta(k- \epsilon a H) (- \epsilon H \dot{a}) \nonumber \\
& \qquad \phantom{space}\times \left( a_{\vec{k},i} \chi_{\vec{k}} e^{-i \vec{k} \cdot \vec{x}}
+ a_{\vec{k},i}^\dagger \chi_{\vec{k}}^\dagger e^{i \vec{k} \cdot \vec{x}}\right).
\label{eq:fi}
\end{align}
Eqs.~\eqref{eq:Langevin} and~\eqref{eq:fi} are straightforward generalizations of the single-field expressions. The multidimensional Fokker-Planck equation associated with Eq.~\eqref{eq:Langevin} governs the evolution of the one-point probability  distribution function $\rho (\bar{\vec{\chi}} ,t)$ describing the probability to observe the field $\chi_{L,i}$ at the value $\bar\chi_{L,i}$ at time $t$; for notational simplicity we omit the overbars henceforth.
From Eqs.~\eqref{eq:wavefns} and~\eqref{eq:fi}, the two-point function for the short-distance quantum noise is
\begin{align}
\nonumber
\langle f_i  (x_1,t_1) f_j (x_2,t_2) \rangle  &=  \frac{H^3}{4\pi^2}\delta_{ij} \delta (t_1-t_2)\\
   &\phantom{sp} \times \frac{\sin \left( \epsilon a H |x_1-x_2|\right)}{\epsilon a H |x_1-x_2|} ,
\end{align}
yielding for the Fokker-Planck equation
\begin{align}
\frac{\partial \rho(\vec{ \chi}_L,t)}{\partial t}
&= \frac{1}{3H}
\left[ \rho(\vec{ \chi}_L,t) \nabla^2 V
+ \vec{\nabla} V \cdot \vec{\nabla} \rho(\vec{ \chi}_L,t)
\right] \nonumber \\
& \hphantom{{}={}} + \frac{H^3}{8 \pi^2} \nabla^2  \rho(\vec{ \chi}_L,t) ,
 \label{eq:fokker_planck}
\end{align}
where the derivatives indicated by $\vec{\nabla}$ are taken with respect to the field space; see also Ref.~\cite{Hardwick:2018sck}, which studied interacting spectator scalars in a general potential.  Here the probability distribution is normalized according to
\begin{align}
\int \ud^N\chi_L \, \rho(\vec{\chi}_L, t) = 1.
\label{eq:norm1}
\end{align}
%\textcolor{red}{This equation was derived in a different scenario in Ref.~\cite{Hardwick:2018sck}, which considered a set of interacting scalar fields without either a global or gauge symmetry.}

To solve for $\rho(\vec{\chi}_L,t)$, we decompose it as a sum of eigenmodes $\Phi$, whose time dependence is described by an eigenvalue $\Lambda$.
The mode functions and their corresponding eigenvalues carry a set of indices $n, m, \ldots$ that are generalizations of the familiar $n,\ell,m$ indices of a 3-dimensional spherical decomposition to $N$-dimensional space.  We make the ansatz
\begin{align}
\rho(\vec{\chi}_L,t)
&= e^{-v(\vec{\chi}_L)}
\sum_{n,m,\dots} a_{n,m,\dots} \Phi_{n,m,\dots}(\vec{\chi}_L) e^{- \Lambda_{n,\dots}(t-t_0)},
\label{eq:rho1}
\end{align}
where $t_0$ is the initial time (at which we impose our initial conditions) and we have defined
\begin{align}
 v(\vec{\chi}_L)
\equiv \frac{4 \pi^2 V(\vec{\chi}_L)}{3 H^4}.
\end{align}
The mode functions are normalized according to
\begin{align}
\int \ud^N \chi_L \, \Phi_{n,m,\dots}(\vec{\chi}_L) \Phi_{n^\prime,m^\prime,\dots}^\dag(\vec{\chi}_L) &= \delta_{n,n^\prime} \delta_{m,m^\prime} \dots,
\end{align}
which is consistent with the normalization condition in Eq.~\eqref{eq:norm1} due to the existence of a mode with zero eigenvalue ($\Lambda_{0} = 0$), which we demonstrate below.
Substituting the ansatz of Eq.~\eqref{eq:rho1} into the Fokker-Planck equation, Eq.~\eqref{eq:fokker_planck}, gives an eigenvalue equation for the modes,
\begin{align}
-  \nabla^2 \Phi_{n,m,\dots}
+  \left[
(\vec{\nabla} v)^2 - \nabla^2 v
\right] \Phi_{n,m,\dots}
&= \frac{8 \pi^2 \Lambda_{n,\dots}}{H^3} \Phi_{n,m,\dots}.
\label{eq:eigenfunction1}
\end{align}

For $\mathrm{SO}(N)$-invariant initial conditions, the probability distribution function depends only on the radial variable $\chi \equiv \sqrt{\vec{\chi}_L \cdot \vec{\chi}_L}$ (see Appendix~\ref{sec:lneq0} for an explicit proof).  The distribution of the radial vev can be obtained from an effective one-dimensional probability density
\begin{equation}\label{eqn:rhoeffeqn}
 \rho_{\mathrm{eff}}(\chi_L,t) \equiv {\chi}_L^{N-1}
 \rho({\chi}_L,t) \cdot \Omega^{(N-1)},
\end{equation}
where $\Omega^{(N-1)}$ is the surface area of a unit $N-1$-sphere, ensuring
\begin{align}
\int_{0}^{\infty} \ud \chi_L \,  \rho_{\mathrm{eff}} ({\chi}_L,t) = 1.
\label{eq:rho1d_norm}
\end{align}

Finally, it is convenient to work with the dimensionless variables
\begin{equation}
  \hat{t}  \equiv  H t, \qquad \hat \chi \equiv \chi_L / H  ,
\end{equation}
and the corresponding probability distribution $\hat{\rho}(\hat{\chi},\hat{t})$, which is normalized according to $\int \ud^{N} \hat{\chi} \, \hat \rho(\hat{\chi},t) = 1 $.
Defining the effective one-dimensional distribution $\hat{\rho}_\mathrm{eff}(\hat{\chi}, \hat{t})$ analogously in terms of $\hat{\rho}(\hat{\chi}, \hat{t})$,
we then seek solutions of the Fokker-Planck equation
\begin{align}
\label{eqn:FPeqnspherical}
%\label{eqn:effPE}
\frac{\partial \hat{\rho}_\mathrm{eff}}{\partial \hat{t}}
&= \frac{1}{8 \pi^2} \left( 2 \frac{\partial^2 v}{\partial\hat\chi^2} + \frac{N-1}{\hat{\chi}^2} \right) \hat{\rho}_\mathrm{eff} \\ \nonumber
& \qquad + \frac{1}{8\pi^2} \left( 2 \frac{\partial v}{\partial \hat\chi} - \frac{N-1}{\hat{\chi}} \right) \frac{\partial \hat \rho_\mathrm{eff} }{\partial \hat{\chi}}
+ \frac{1}{8\pi^2} \frac{\partial^2 \hat{\rho}_\mathrm{eff}}{\partial \hat{\chi}^2}
\end{align}
for $\hat \rho_\mathrm{eff}(\hat{\chi}, \hat{t})$.

%%%%%%%
\subsubsection{Zero modes for $\mathrm{SO}(N)$-invariant potentials}
%%%%%%%

The existence of a zero mode can be demonstrated analytically for an arbitrary stable $\mathrm{SO}(N)$-invariant potential.  Note that
\begin{align}
\Phi_0(\vec{\chi}_L) &= \mathcal{N} e^{-v(\vec{\chi}_L)}
\end{align}
satisfies Eq.~\eqref{eq:eigenfunction1} with eigenvalue $\Lambda_0 = 0$.

Because all eigenvalues are non-negative, the zero mode determines the late-time asymptotic form of the probability distribution.  Therefore, the coefficient $a_0$ in the expansion in Eq.~\eqref{eq:rho1} is not sensitive to the initial condition at $t=t_0$. Instead, $a_0$ can be determined by requiring the probability distribution function to be unit-normalized in the asymptotic future, $\lim_{t \rightarrow \infty} \int \ud^N \chi_L \, \rho(\chi_L) = 1$.  Recognizing that the integrand is proportional to $\Phi_0(\vec{\chi}_L)^2$ and using the eigenmode normalization condition, we see that $a_0 = \mathcal{N}$, giving
\begin{align}\label{eqn:eqdist}
\lim_{t \rightarrow \infty} \rho(\vec{\chi}_L)= \mathcal{N}^2 e^{- 2 v(\vec{\chi}_L)},
\end{align}
where the constant $\mathcal{N}$ depends on both the dimensionality of field space and the functional form of the potential.

%%%%%%%%%%%%%%%%
\subsection{SU(2) and other global symmetry groups}
%%%%%%%%%%%%%%%%

To generalize this treatment to other compact symmetry groups, we begin with an instructive example. Consider  a scalar field  transforming as the fundamental representation of a global $\mathrm{SU}(2)$ symmetry, which can be parametrized as
\begin{align}
\chi = \begin{pmatrix}
\chi_1 + i \chi_2 \\
\chi_3 + i \chi_4
\end{pmatrix}.
\end{align}
We assume the action is invariant under $\mathrm{SU}(2)$ rotations, so the potential is a function of $\chi^\dagger \chi = \chi_1^2 + \chi_2^2 + \chi_3^2 + \chi_4^2$. To derive the Fokker-Planck equation describing the evolution of the vev, we  proceed  as above for the $\mathrm{SO}(N)$ case,  decomposing the fields into long- and short-wavelength components and using the short-wavelength free-field correlation function to obtain the corresponding stochastic diffusion.  Since we are interested in the magnitude of the vev and not its direction in field space, it is again convenient to work in spherical coordinates on field space. Explicitly, we write the scalar field as
\begin{align}
\label{eq:chispinor}
\chi_L = \chi \begin{pmatrix}
\cos\theta_1 + i \sin\theta_1 \cos\theta_2 \\
\sin\theta_1 \sin\theta_2 \cos\phi + i \sin\theta_1\sin\theta_2 \sin\phi
\end{pmatrix},
\end{align}
where
$\theta_1,\theta_2 \in [0,\pi]$ and $\phi \in [0,2\pi)$. The space where the vacuum expectation value takes values, $\mathbb{C}^2$, is isomorphic to $\mathbb{R}^4$, and the Laplacian over the (rescaled) field variables in the parametrization of Eq.~\eqref{eq:chispinor} is
\begin{align}
\hat{\nabla}^2
&= \frac{\partial^2}{\partial \hat{\chi}^2} + \frac{3}{\hat{\chi}} \frac{\partial}{\partial \hat{\chi}} + \frac{1}{\hat{\chi}^{2}} \nabla^2_{S^{3}},
\end{align}
where
\begin{align}
& \nabla_{S^{3}}^2  = \frac{1}{\sin^2\theta_2} \frac{\partial}{\partial \theta_2} \left( \sin^2\theta_2 \frac{\partial }{\partial \theta_2} \right)  \\\nonumber
& + \frac{1}{\sin^2\theta_2} \left( \frac{1}{\sin\theta_1} \frac{\partial}{\partial \theta_1} \left( \sin\theta_1 \frac{\partial }{\partial \theta_1} \right) + \frac{1}{\sin^2\theta_1} \frac{\partial^2 }{\partial \phi^2} \right) .
\end{align}
The results of the previous section can now be applied straightforwardly; for example, at late times the distribution function takes its equilibrium form,
\begin{align}
\lim_{t \rightarrow \infty} \rho(\vec{\chi}_L) = \mathcal{N}^2 e^{-2 v(\chi_L)},
\end{align}
where again the precise value of the constant $\mathcal{N}$ depends on the dimensionality of field space.

This SU(2) example makes it clear that the derivation of the Fokker-Planck equation depends only on the manifold structure of the field space of the spectator scalar and not on the manifold structure of the group itself.
The probability distribution function (for linear representations) is  therefore determined by the number of real degrees of freedom possessed by the scalar, as well as the potential. We expect that our analysis holds for a broad range of possible scalar field theories; however, it would not apply directly to theories that have a nontrivial field space metric or topology.

%%%%%%%%%%%%%%%%%%%%%%%%%%%%%%%%%%%
\subsection{Example: $N$-dimensional evolution in a quartic potential}\label{sec:quartic}
%%%%%%%%%%%%%%%%%%%%%%%%%%%%%%%%%%%

To illustrate the effects of the dimensionality of the field space, we consider an example where an $N$-dimensional scalar field moves in a $\mathrm{SO}(N)$-symmetric quartic potential.  We focus on three effects: first, the equilibrium distribution and vev attained by the field depend on the dimensionality of the field space; secondly, in larger dimensionality spaces, the probability distribution approaches its equilibrium value faster; and thirdly, we discuss the evolution of the resulting large-$\chi$ tail of the probability distribution.  The advantages of this quartic example are that it allows analytic solutions and it helps to develop some quantitative intuition for the $N$-dependence of the results.

%%%%%%%%
\subsubsection{Equilibrium distribution and vev}
%%%%%%%%

The coefficient $\mathcal{N}$ in Eq.~\eqref{eqn:eqdist} can be determined analytically for the quartic potential, which we write as
\begin{align}\label{eqn:quarticpot}
V = \frac{\lambda}{4} (\vec{\chi} \cdot \vec{\chi})^2\equiv \frac{\lambda}{4} \chi^4 ,
\end{align}
where the normalization of the coupling constant has been chosen to agree with
%With this choice, the results of
the conventions of Ref.~\cite{Starobinsky:1986fx,Starobinsky:1994bd} for $N=1$,
facilitating comparison.
For this potential, we find the asymptotic probability distribution is given by
\begin{align}
\lim_{\hat t \rightarrow \infty} \hat{\rho}(\hat{\chi}, \hat{t})
&= \frac{2 \Gamma(N / 2)}{\Gamma(N / 4)}  \left( \frac{2}{3} \right)^{N / 4} \lambda^{N/4}
e^{-2v(\hat{\chi})},
\label{eq:asymp}
\end{align}
which reduces to the results of Ref.~\cite{Starobinsky:1986fx} for $N=1$.
Equivalently, the asymptotic one-dimensional probability density over the rescaled field is
\begin{align}
\label{eq:asymp1d}
    \lim_{\hat{t} \rightarrow \infty} \hat{\rho}_{\mathrm{eff}}(\hat{\chi}, \hat{t})
    = \frac{4 \lambda^{N/4} }{\Gamma(N / 4)} \left( \frac{2 \pi^2}{3} \right)^{N / 4} \hat{\chi}^{N-1} e^{-2 \pi^2 \lambda \hat{\chi}^4 / 3}.
\end{align}

From Eq.~\eqref{eq:asymp1d} we find that the equilibrium average vev for a spectator field in a quartic potential is
\begin{align}
\left< \hat\chi^2 \right>
&= \lambda^{-1/2}  \sqrt{ \frac{3}{2}} \frac{\Gamma((2 + N) / 4)}{\pi \Gamma(N / 4)},
\label{eq:vev}
\end{align}
which increases as $\sqrt{N}$ at large $N$. 

This dependence on $N$ is a consequence of the enlarged field space available for the random walk of the spectator vev in a quartic potential: for a one-dimensional scalar field, each stochastic jump is either towards or away from the origin, while in a multi-dimensional space, there are many independent directions into which the vev can diffuse.

As for the one-dimensional case~\cite{Starobinsky:1994bd}, higher mode functions can be found numerically.  We present our results for the higher eigenmodes in Appendix~\ref{sec:lneq0}.  Critically, we find that for any fixed mode (e.g., the first excited mode) the corresponding eigenvalue increases as the dimensionality $N$ increases.  Eq.~\eqref{eq:rho1} then implies that the probability distribution approaches its equilibrium value faster at larger $N$, as we now discuss.

%%%%%%%%%
\subsubsection{Approach to equilibrium}\label{sec:eqapproach}
%%%%%%%%%

The eigenvalues of excited modes in the probability distribution govern the evolution of $P(\hat \chi)$ to its asymptotic form via the exponential factor $e^{- \Lambda_{n,m,\dots} (t-t_0)} $ in Eq.~\eqref{eq:rho1}.  In Appendix~\ref{sec:lneq0}, we show that for a quartic potential with fixed coupling $\lambda$ the eigenvalue for a given mode increases with $N$.  Therefore, any result derived from the approach to equilibrium depends on the number of degrees of freedom in the scalar sector.  In particular,  the asymptotic probability distribution and equilibrium vev derived above become applicable earlier (i.e., after fewer $e$-foldings) for scenarios with more degrees of freedom.

In order to study the approach to equilibrium, we evolve Eq.~\eqref{eqn:FPeqnspherical} numerically. We take Gaussian initial conditions
\begin{equation}\label{eqn:IC}
\hat \rho(\hat{\chi},0) \propto e^{-  (\hat \chi / \sigma)^2},
\end{equation}
with $\sigma \approx 0.0563$ (using Eq.~\eqref{eqn:initial-condition-width}), and consider the subsequent evolution of the probability distribution.

The approach to the equilibrium distribution depends  on both the quartic coupling $\lambda$ and the dimensionality of the field $N$. The qualitative dependence on $\lambda$ is straightforwardly determined by noting that the parameter $\lambda$ can be removed from the dynamics by rescaling the field according to $\hat{\chi} \to \lambda^{1/4} \hat{\chi}$ (see Appendix~\ref{sec:lneq0} for details). However, after rescaling the field, $\lambda$ reappears in the initial condition, Eq.~\eqref{eqn:IC}.
Lower $\lambda$ then corresponds to a more sharply peaked initial condition with otherwise identical evolution,  therefore taking longer to reach equilibrium. In what follows, to explore the dependence on $N$ we fix $\lambda = 0.05$.

\begin{figure}
\begin{center}
\includegraphics[width=\columnwidth]{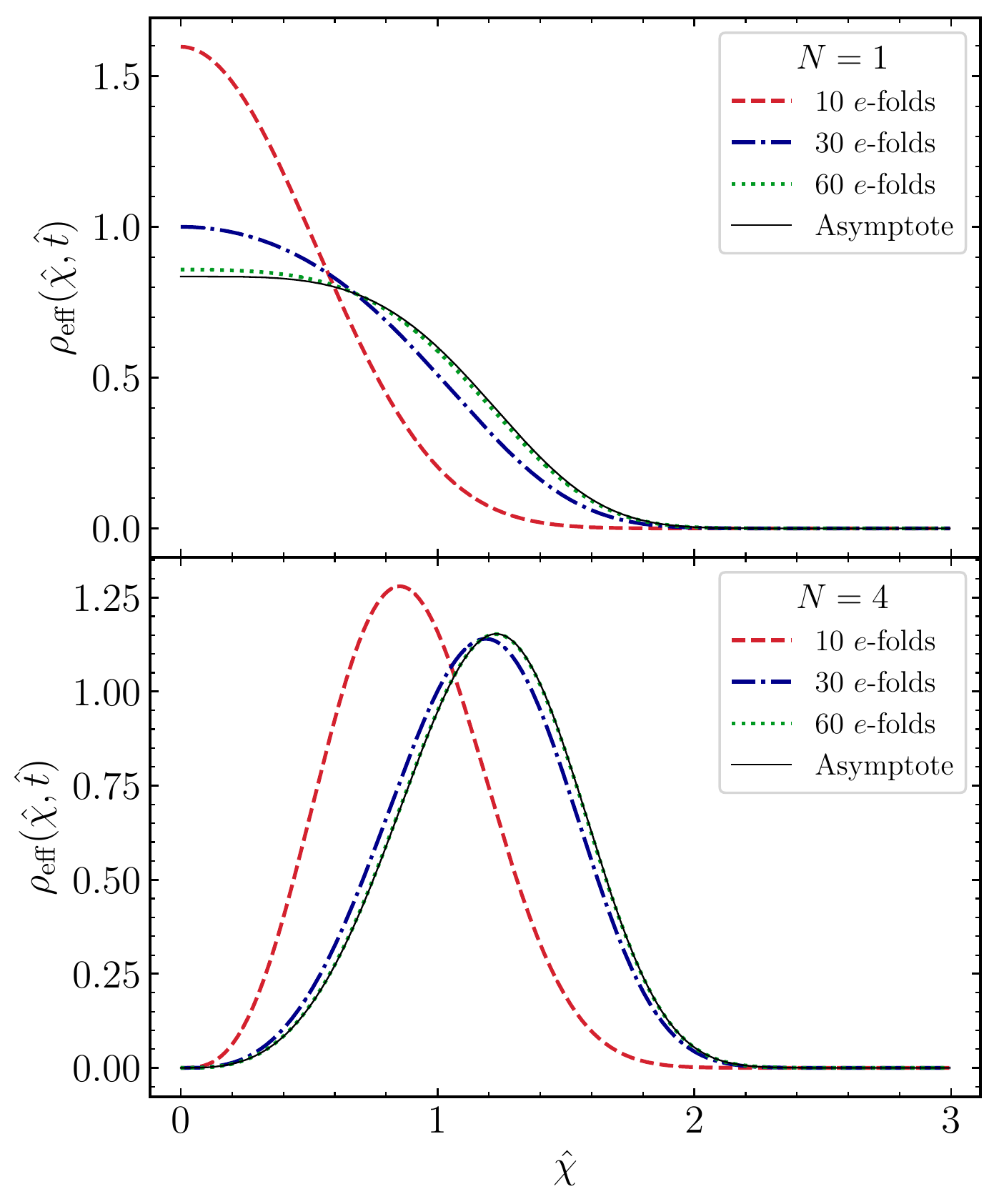}
\caption{The effective probability distribution $\hat \rho_{\mathrm{eff}}(\hat{\chi},\hat t)$ as a function of $\hat{\chi} = \chi / H$, for a quartic potential with $\lambda = 0.05$ and $N=1$ (top) or $N=4$ (bottom).}
\label{fig:approach_asymptotic}
\end{center}
\end{figure}

The evolution of the effective one-dimensional probability with the number of $e$-folds is shown in Fig.~\ref{fig:approach_asymptotic}, for $N=1$ (top) and $N=4$ (bottom).  Comparing the curves for 30 and 60 $e$-folds to the asymptotic result demonstrates that the $N=4$ scenario relaxes to equilibrium faster.

%%%%%%%%%%%%%
%%%%%%%%%%%%

\begin{figure}
\begin{center}
\includegraphics[width=\columnwidth]{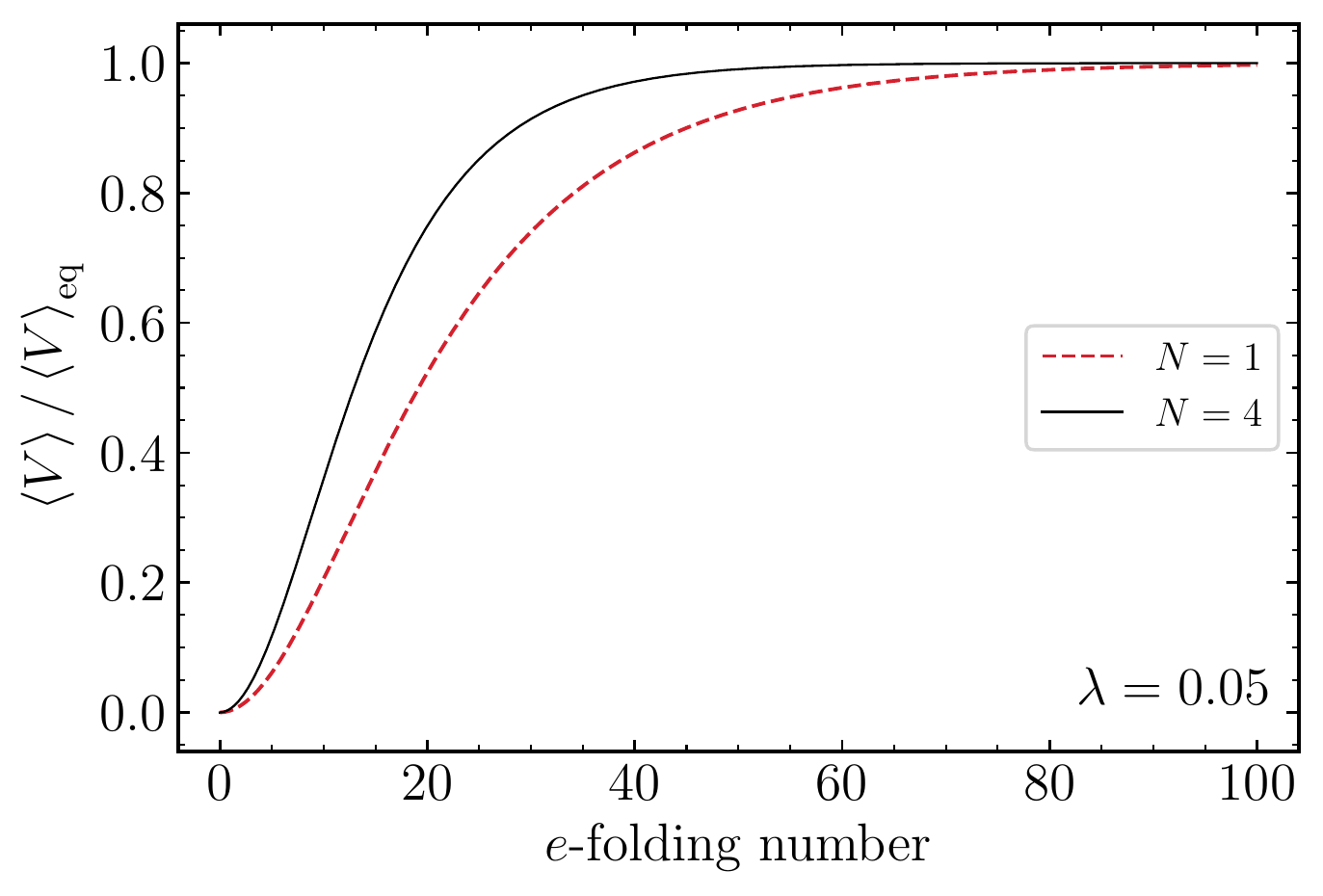}
\caption{Evolution of the ratio of the expectation value of the energy density $V = \lambda \chi^4 / 4 $ to its equilibrium value as a function of time (number of $e$-folds), for scalar fields with $N=1$ and $N=4$ degrees of freedom and fixed $\lambda = 0.05$.
\label{fig:Eq_Approach}
}
\end{center}
\end{figure}

To illustrate the impact of faster relaxation to equilibrium in higher-dimensional field spaces, we consider the evolution of the energy density $\left< V \right> = \lambda \left< \chi^4 \right> / 4$.  In a larger field space the equilibrium vacuum expectation value is enhanced, which corresponds to an enhanced energy density stored in the spectator scalar field.  For a field with $N$ real degrees of freedom in a quartic potential, the asymptotic probability distribution gives
\begin{align}
\left< V \right>_{\rm eq} = \frac{\lambda}{4} \left< \chi^4 \right>_{\rm eq} = \frac{3 N H^4}{32 \pi^2}.
\end{align}
Because the equilibrium value is linearly dependent on the number of scalar degrees of freedom, to isolate the effect of the speed of relaxation as the dimensionality is varied we consider the evolution of $\left< V \right> / \left< V \right>_{\rm eq}$.  This quantity is shown in Fig.~\ref{fig:Eq_Approach} as a function of time, measured by the number of $e$-folds of expansion.  As expected from the eigenvalue analysis, the scalar field with four degrees of freedom approaches its equilibrium value significantly faster than the singlet scalar field.
After 60 $e$-folds of inflation, the $N=4$ scalar field energy density is 99.7\% of its equilibrium value while the $N=1$ scalar field energy density is 96.2\% of its equilibrium density.
%%%%%%%%%%%%
%%%%%%%%%%%%

%%%%%%%%
\subsubsection{Effects of higher-dimensional field space on the tail of the distribution}
%%%%%%%%

The high-scale instability in the SM Higgs potential implies that any Hubble patch where the Higgs field probes sufficiently far beyond the instability scale evolves to the true vacuum.  Therefore, any enhancement of the probability for the Higgs to reach large field values may have critical consequences for the safety of our own Universe.
We have seen above that a higher-dimensional field space leads to significant enhancement in the equilibrium vev attained by the field. We now consider the effect of the higher dimensionality of the field space on the tail of the distribution at large values of the vev.  Although we treat a purely quartic potential with a global symmetry in this section, the quantitative intuition we develop here is useful for the more complicated case of the SM Higgs we consider below in Sec.~\ref{sec:inflation}.

\begin{figure}
\begin{center}
\includegraphics[width=\columnwidth]{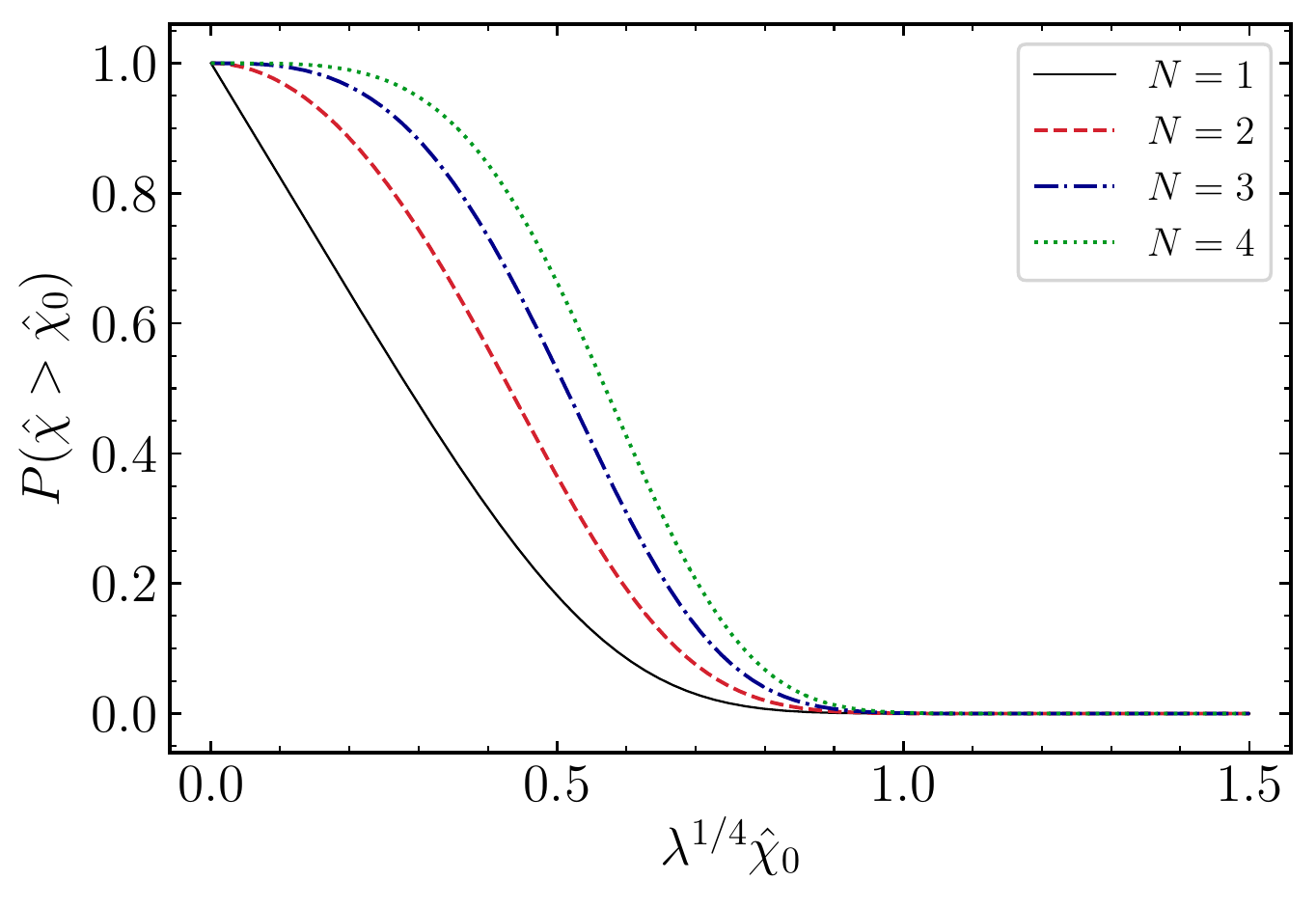}
\caption{Asymptotic values of $P(\hat{\chi} > \hat{\chi}_0)$ for a quartic potential, as a function of $\lambda^{1 / 4} \hat{\chi}_0$, for scalar fields with $N=1$ to $N=4$ real degrees of freedom.}
\label{fig:Probability_chi_large}
\end{center}
\end{figure}

\begin{figure}
\begin{center}
\includegraphics[width=\columnwidth]{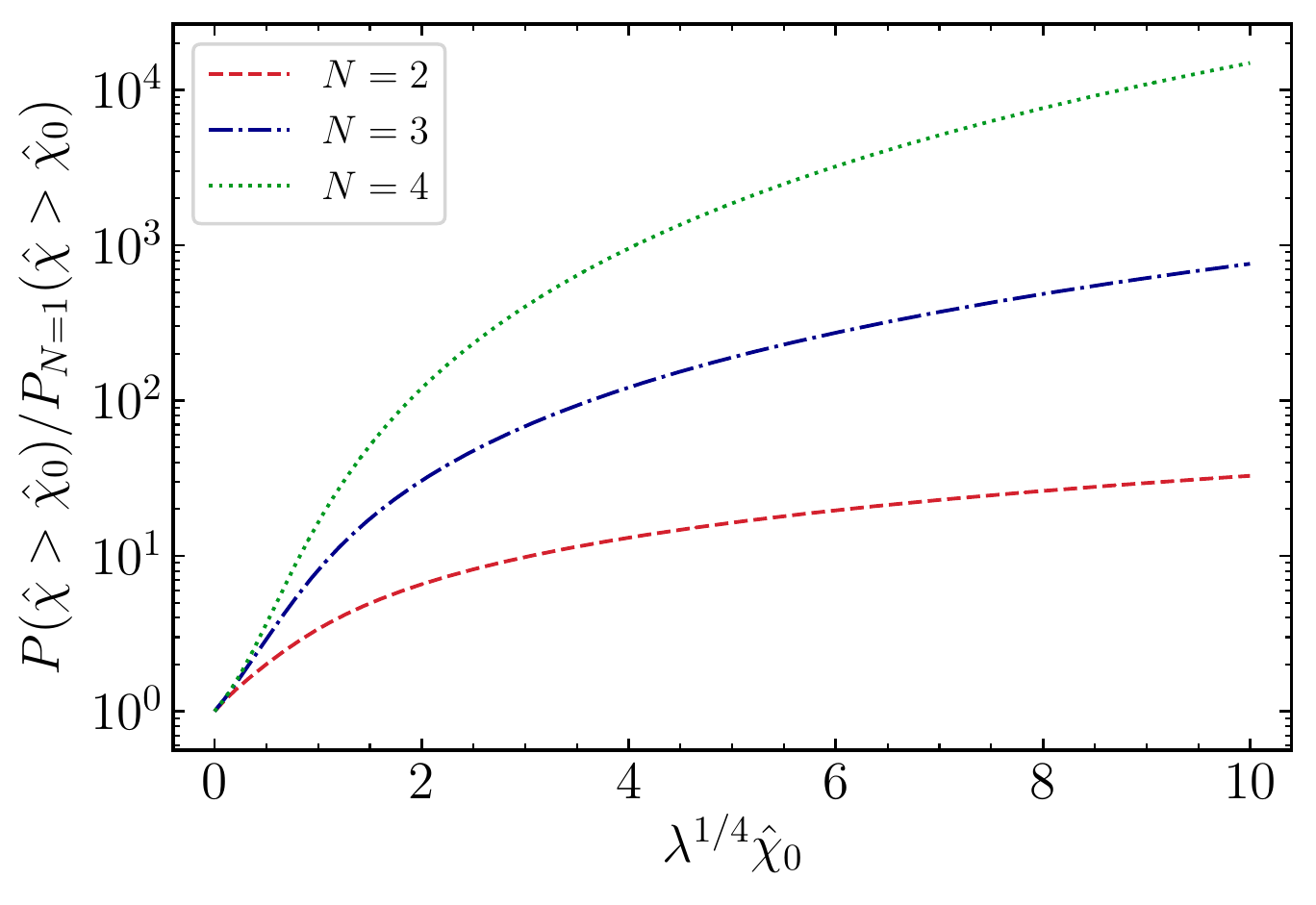}
\caption{Logarithm of the ratio $P(\hat{\chi} > \hat{\chi}_0)$ for $N$ real degrees of freedom to $N=1$  in a quartic potential, as a function of $\lambda^{-1 / 4}\hat{\chi}_0$.  The probability of finding $\hat{\chi} > \hat{\chi}_0$ can be enhanced by several orders of magnitude at large $\lambda^{-1 / 4} \hat{\chi}_0$.}
\label{fig:tail}
\end{center}
\end{figure}

At late times, when the probability distribution approaches its asymptotic form, the probability that the vev takes values larger than any given field value $\hat{\chi}_0\equiv \chi_0/H$ is
\begin{align}
P(\hat{\chi} \geq \hat{\chi}_0) &= 1 - \int_0^{\hat{\chi}_0} \ud\hat{\chi} \,  \rho_{\mathrm{eff}}(\hat{\chi}), \nonumber \\
\label{eq:p4analytic}
&= \frac{\Gamma(N / 4, 2 \pi^2 \lambda \hat{\chi}_0^4 / 3)}{\Gamma( N / 4)},
\end{align}
where $\Gamma(a,z)$ is the incomplete Gamma function.  The dependence of $P(\hat \chi \geq \hat{\chi}_0)$ on the number of real degrees of freedom $N$ is shown in Fig.~\ref{fig:Probability_chi_large}.  As $N$ increases, there is more support at large field values, and therefore the probability of finding $\hat{\chi} \geq \hat{\chi}_0$ falls off less rapidly.  The tail of the probability distribution is thus enhanced by several orders of magnitude, as shown in Fig.~\ref{fig:tail}.

%%%%%%%%%%%%%

%%%%%%%%%%%%%%%%

The current observable Universe is comprised of approximately $e^{3n}$ patches of size $\sim H$, where $n\gtrsim 60$ is the number of $e$-folds of inflation. The probability that no patch in our past light cone has a value of $\hat{\chi} $ greater than some value $\hat{\chi}_{-180}$ is found by solving
\begin{align}
\label{eq:chi_-180_def}
P(\hat{\chi} > \hat{\chi}_{-180}) \lesssim e^{-180} .
\end{align}
Solving this equation  for $ \hat{\chi}_{-180}$ with $N=1$ in a  quartic potential gives
\begin{align}
\label{eq:chicr1}
\lambda^{1/4} \hat{\chi}_{-180,N=1}> 2.27
\end{align}
using the equilibrium probability distribution of Eq.~\eqref{eq:p4analytic}.
Because the large-$\chi$ tail of the probability distribution is enhanced as $N$ increases, solving the same equation for $N>1$ results in larger values of $\hat{\chi}_\mathrm{-180}$.  However, the resulting numerical shift in $\hat\chi_\mathrm{-180}$ is small: although the tail of the probability distribution is enhanced by several orders of magnitude as $N$ increases (see Fig.~\ref{fig:tail}), this enhancement is counterbalanced by the exponential decrease of the probability distribution with $\hat{\chi}$ (independent of the dimensionality).

Given some $\hat\chi_{0,N=1}$, we can analytically  solve the equation
\begin{align}
P(\hat{\chi} > \hat{\chi}_{0,N=4})|_{N=4} = P(\hat{\chi} > \hat{\chi}_{0,N=1})|_{N=1}
\end{align}
for $\hat{\chi}_{0,N=4}$ in a quartic potential to obtain the value $\hat{\chi}_{0,N=4}$ for which the probability in the tail of the four-dimensional distribution is equal to the specified probability in the tail of the one-dimensional distribution.  We find
\begin{align}
\hat{\chi}_{0,N=4} = \left[ \frac{3}{2 \pi^2 \lambda} \ln \left( \frac{\Gamma(1 / 4)}{\Gamma(1 / 4,2 \pi^2 \lambda \hat{\chi}_{0,N=1}^4 / 3)} \right) \right]^{1 / 4}.
\label{eq:relate_chi_crit}
\end{align}

Using Eq.~\eqref{eq:relate_chi_crit}, we find that $\lambda^{1/4} \hat{\chi}_{-180,N=1} \gtrsim 2.27 $ corresponds to
\begin{align}
\label{eq:chicr4}
\lambda^{1/4} \hat{\chi}_{-180,N=4} \gtrsim 2.29.
\end{align}

This result holds for the equilibrium probability distribution.  However, as we have noted above, the approach to the equilibrium distribution is faster for $N>1$ than for $N=1$, which can be important if the inflationary epoch is too short for the probability distribution to obtain its asymptotic form.  For sufficiently narrow initial conditions, the value of $\chi_{-180}$ can be substantially smaller than the asymptotic results when inflation ends before the equilibrium distribution is reached.

\begin{figure}
\begin{center}
\includegraphics[width=\columnwidth]{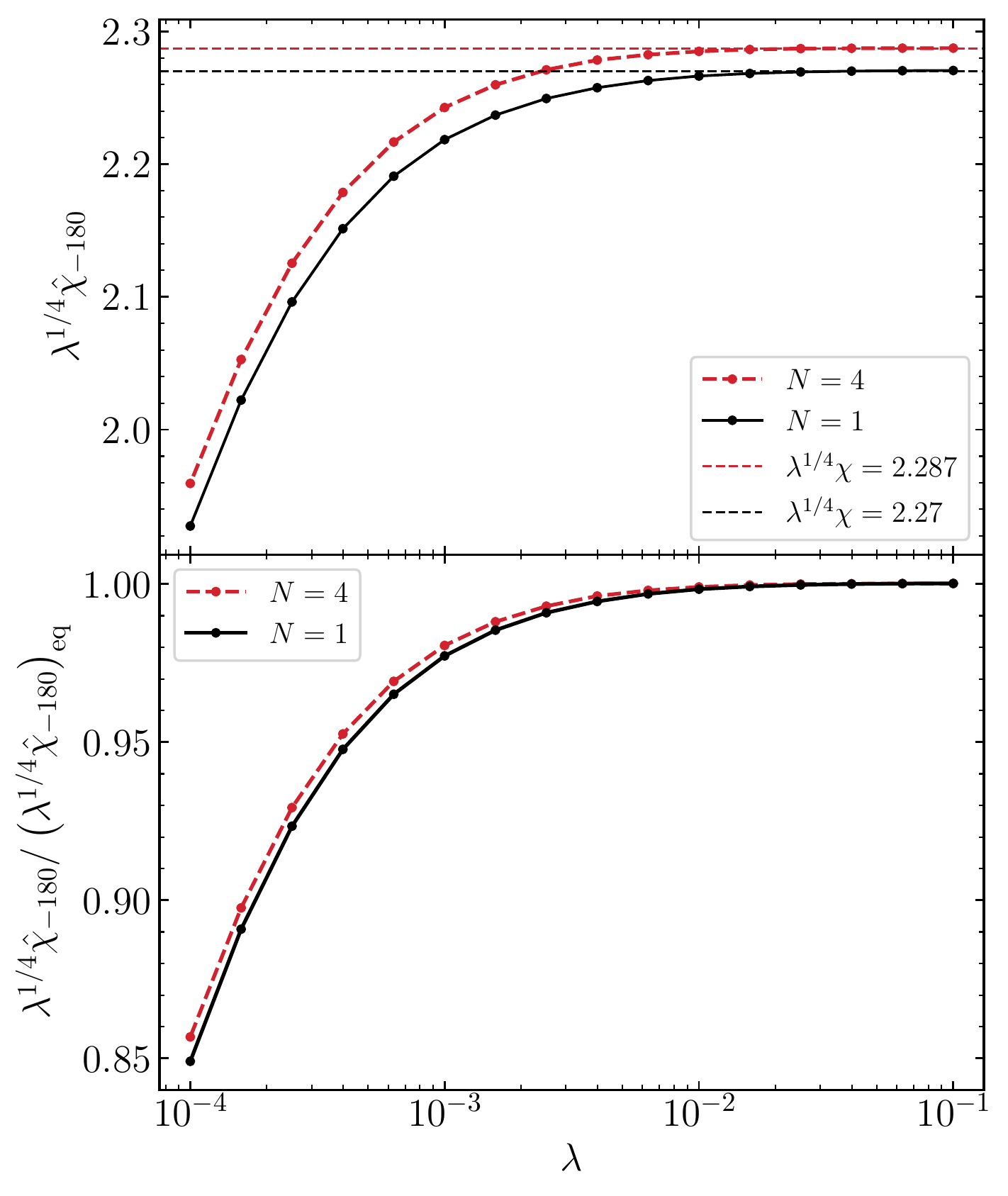}
\caption{Top: $\lambda^{1 / 4} \hat{\chi}_{-180}$, with $\hat{\chi}_{-180}$ defined in Eq.~\eqref{eq:chi_-180_def}.  Dots indicate the value after 60 $e$-folds while  lines indicate the values obtained from the asymptotic probability distribution function, Eqs.~\eqref{eq:chicr1} and~\eqref{eq:chicr4}.  Bottom: The ratio of $\lambda^{1 / 4} \hat{\chi}_{-180}$ to its equilibrium (asymptotic) value.}
\label{fig:chi_-180}
\end{center}
\end{figure}

We study the tail of the distribution  by evolving the Fokker-Plank equation numerically, as described in Appendix \ref{sec:comp}. After evolving for 60 $e$-folds, we  determine $\hat{\chi}_{-180}$, with results shown in Fig.~\ref{fig:chi_-180}. As expected, we see that at sufficiently small couplings ($\lambda \lesssim 10^{-2.5}$) the field value $\hat{\chi}_{-180}$ is significantly smaller than its asymptotic value.  For $\lambda \sim 10^{-4}$, we have $\lambda^{1 / 4} \hat{\chi}_{-180,N=1} \approx 1.92$ and $\lambda^{1 / 4} \hat{\chi}_{-180,N=4} \approx 1.96$.
The bottom panel of Fig.~\ref{fig:chi_-180} shows the ratio of $\lambda^{1 / 4} \hat{\chi}_{-180,N=1} $ to its equilibrium value. We see that while $N=4$ does approach its equilibrium value at slightly smaller values of $\lambda$, as expected from Sec.~\ref{sec:eqapproach}, the difference between $N=1$ and $N=4$ lines remains small over four decades in $\lambda$, never differing by more than 2 percent.

Given the significant effect of the dimensionality of the field space on the vev, energy density, and the speed of approach to equilibrium, it may appear surprising that the probability to obtain very large vevs is not similarly affected. However, some insight into the large-vev behavior can be obtained by considering the
the Fokker-Planck equation for $\hat{\rho}_{\mathrm{eff}}$, Eq.~\eqref{eqn:FPeqnspherical}.
At large field values $\hat{\chi} \gg 1$, the terms involving $N$ are unimportant compared to the potential (assuming that $\partial v/\partial \hat\chi$ falls slower than $1/\hat{\chi}$). That is, the dynamics far from the origin are dominated by the forcing due to the potential gradients. For the quartic potential it is straightforward to see that, for field values much larger than the equilibrium vev, the dynamics due to the extra degrees of freedom become subdominant.

%
%%%%%%%%%%%%%%%%%%%%%%
%%%%%%%%%%%%%%%%%%%%%%%%

%%%%%%%%%%%%%%%%%%%%%%%%%%%%%%%%%%%
\section{Gauging the Symmetry}
\label{sec:gauging}
%%%%%%%%%%%%%%%%%%%%%%%%%%%%%%%%%%%

So far, we have considered fields transforming under a global symmetry.
In this section, we gauge the symmetry and introduce the associated gauge fields.
When the scalar field obtains a vacuum expectation value during its random walk, the gauge symmetry is spontaneously broken, and the physical spectrum develops a mass gap parametrically given by $m\sim g\chi$, where $g$ is the gauge coupling. The physical content of a spontaneously broken gauge theory is often made more explicit by going to unitary gauge, where Goldstone bosons are no longer present in the theory as propagating fields.\footnote{For a discussion of unitary gauge at a general location in field space, where $\partial V/\partial \chi_i\neq 0$, see~\cite{Mooij:2011fi}.}
However, the bad UV behavior of the longitudinal polarization of the gauge-boson propagator in unitary gauge introduces spurious divergences in loop diagrams. In Minkowski space, these divergences require a new counterterm to cancel them~\cite{Weinberg:1973ew,  GrosseKnetter:1992nn, Hertzberg:2018kyi}:
\beq
\label{eq:ct}
\mathcal{L}_{ct} = -(N-1) i \delta^4 (0) \ln \left(\chi \right),
\eeq
where $N-1$ is the number of massive gauge bosons, or equivalently the number of eaten Goldstone bosons.
The appearance of this counterterm in unitary gauge can also be seen starting from the Fadeev-Popov Lagrangian in path integral calculations~\cite{Lee:1973fn, GrosseKnetter:1992nn}. For simplicity we take all Goldstone bosons to be eaten, as in the SM, but the discussion should generalize straightforwardly to more general patterns of spontaneous breaking of linearly-realized symmetries.

\begin{figure}
\begin{center}
\includegraphics[width=0.23\columnwidth]{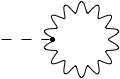}
\phantom{space44}
\includegraphics[width=0.60\columnwidth]{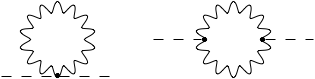}
\caption{Example one-loop diagrams with one and two external Higgs bosons that give rise to spurious quartic divergences in unitary gauge. }
\label{fig:fdiags}
\end{center}
\end{figure}

Derivations of the unitary gauge counterterm from explicit loop calculations~\cite{GrosseKnetter:1992nn,  Hertzberg:2018kyi}
clarify how the divergent $\delta^4(0)$ in Eq.~\eqref{eq:ct}  should be understood.  The spurious quartic divergences that appear in unitary gauge come from one-loop diagrams of the kind shown in Fig.~\ref{fig:fdiags}.
%%%%%%%%%%%%%%%%%%%%%%
%%%%%%%%%%%%%%%%%%%%%%%%
In these diagrams, the unitary gauge form of the gauge boson propagator gives rise to momentum integrals of the form
\begin{align}
    \label{eq:badloops}
    \int \frac{\ud^4 k}{(2\pi)^4} \frac{k^2}{k^2 - m^2} .
\end{align}
The natural regularization of this momentum integral yields $\delta^4 (0) = \int \ud^4 k/ (2\pi)^4 \equiv i \Lambda^4$, where $\Lambda$ is a UV cutoff in $k$-space, and the factor of $i$ arises from the standard Wick rotation employed to evaluate four-momentum integrals.  With this regularization prescription,
the Lagrangian of Eq.~\eqref{eq:ct} is manifestly Hermitian.  These quartic divergences appear in correlation functions with any number $n$ of external Higgs bosons, in such a way that the counterterms needed at each individual $n$ can be resummed to yield the logarithmic expression in Eq.~\eqref{eq:ct}.

Now let us consider what the consequences of these spurious quartic divergences are in de Sitter space. In de Sitter space, the UV structure of the unitary gauge theory is identical to that in Minkowski space. We therefore expect the de Sitter computations that lead to the analogue of Eq.~\eqref{eq:badloops} to generate similar UV divergences. However, we also expect a number of new features in de Sitter due to the presence of the de Sitter horizon which alters the form of the propagator in the IR. Because modes with masses $m \lesssim H$ behave differently than those with $m \gtrsim H$, we expect qualitatively different effects in these two regimes.

In constructing the Langevin and Fokker-Planck equations for the long-wavelength modes above, the split between short and long wavelength modes  imposes an effective IR cutoff on the theory at the horizon scale. The effect of this cutoff depends on the mass of the gauge bosons. On the one hand, for gauge boson masses $m \ll H$, the IR behavior of the unitary gauge contribution to the divergent momentum integrals is controlled by the IR cutoff.  In this case, after the UV divergences in the one-loop effective action are cancelled by the counterterm of Eq.~\eqref{eq:ct}, the loop integrals of Eq.~\eqref{eq:badloops} produce a finite and physical contribution of the form
\beq
\delta \mathcal{L} = (N-1) \Lambda_{IR}^4 \ln \left(\chi \right) = (N-1) \frac{3 H^4}{8 \pi^2} \ln \left(\chi \right),
\eeq
where in the second equality we took the fourth power of the IR cutoff to be the inverse of the subhorizon four-volume.  This subhorizon volume is the product of one Hubble time and the volume of a spatial three-sphere,
\beq
V_{hor} = \frac{4}{3} \pi \left(\frac{1}{H}\right)^3 \times \frac{2\pi}{H}.
\eeq
On the other hand, when the gauge boson mass is above the Hubble scale, $m \gtrsim H$, it is instead the gauge boson mass that controls the IR behavior. In this case, the pole at the physical gauge boson mass is within the region of integration in Eq.~\eqref{eq:badloops}.  Accordingly, the IR cut-off from the finite number of modes within the Hubble volume results in a negligible contributions to the correlation functions.  This reflects the mass gap in the physical spectrum of the spontaneously broken gauge theory: when the mass gap is larger than the Hubble scale, only the radial mode remains light, and the field is effectively one-dimensional.

To summarize, when the gauge boson mass is below the Hubble scale, in de Sitter space the bad UV behavior of the gauge boson propagators in unitary gauge induces a contribution to the Langevin equation for the long-wavelength fluctuations. This contribution acts as an addition to the potential for the radial mode
\begin{eqnarray}
\label{eq:vug}
V(\chi)
& = & \frac{\lambda}{4}\chi^4  -   (N-1) \frac{3 H^4}{8 \pi^2} \ln \left(\chi \right)\\\nonumber
&  \equiv & V_0 (\chi) + V_{ug} (\chi) .
\end{eqnarray}
The Fokker-Planck equation for a single real field in unitary gauge is then
%
%\beq
\begin{eqnarray}
\label{eq:1dug}
    \frac{8\pi^2}{H^3} \ddt{\rho}
    & = & \frac{8\pi^2}{3 H^4} \left( \ddxx{V_0} + \frac{(N-1)}{\chi^2} \right) \rho \\
\nonumber
&&+ \frac{8\pi^2}{3 H^4} \left(\ddx{V_0} -  \frac{(N-1)}{\chi}  \right) \ddx{\rho} + \ddxx{\rho}.
\end{eqnarray}
%\eeq
%
However, this is exactly the same as the Fokker-Planck equation  describing the evolution of the radial mode in an $N$-dimensional field space subject to the potential $V_0$, as we now show.  Taking $\rho$ to depend only on $\chi$, i.e,  isotropic in field space, the $N$-dimensional Fokker-Planck equation is
\begin{eqnarray}
\nonumber
\frac{8\pi^2}{H^3} \ddt{\rho} &=& \frac{8\pi^2}{3 H^4}\left[ \rho \frac{1}{\chi^{N-1}} \ddx{} \left( \chi^{N-1}\ddx{V_0}\right)+\ddx{V_0}\ddx{\rho}  \right]\\
\label{eq:multid}
&& + \frac{1}{\chi^{N-1}}\ddx{}\left(\chi^{N-1}\ddx{\rho}\right).
\end{eqnarray}
But when Eq.~\eqref{eq:multid} is written in terms of the effective one-dimensional probability distribution for the radial mode $\rhoeff$ (defined in Eq.~\eqref{eqn:rhoeffeqn}), it precisely matches Eq.~\eqref{eq:1dug} with the replacement $\rho\to \rho_{\rm eff}$.

Since the vev of the scalar field controls the size of the mass gap, at sufficiently large values of the vev ($\chi\gtrsim H/g$) the field necessarily appears one-dimensional, as only the radial mode remains light compared to the Hubble scale.  Thus we generically expect a scalar field transforming non-trivially under a gauge symmetry to transition between an $N$-dimensional regime at small vevs to a one-dimensional regime at large vevs.  To fully describe this transition would require a full loop calculation in de Sitter space, which is beyond the scope of this paper; however, as we show below, this treatment is already sufficient to obtain interesting results for the SM Higgs.

Gauging the symmetry also introduces a distinct new correction to the potential from the  backreaction of the associated gauge bosons.
In unitary gauge, the gauge fields that correspond to spontaneously broken generators acquire mass and a longitudinal degree of freedom via the Higgs mechanism. This mass term breaks the conformal symmetry, inducing particle production of the massive degrees of freedom in the de Sitter background. This process results in a physical bath of gauge bosons on the horizon scale that can then backreact on the radial mode.

In the Hartree approximation, we can estimate the correction to the scalar potential from the interaction with this background of gauge bosons as
\begin{align}
\Delta V &\approx \frac{1}{2} g^2  \left< A^\mu A_\mu \right> \chi^2.
\end{align}
Approximating the gauge boson expectation value by $\left< A^\mu A_\mu \right>  \sim 3 H^2/(4\pi)^2$,
we see that the physical bath of gauge bosons induces a small correction to the scalar mass,
\begin{align}
\delta m^2_\chi \sim 3 g^2\frac{H^2}{(4\pi)^2}.
\end{align}
For $g \lesssim 1$, this correction is unimportant. Further, as the scalar wanders further from the origin and the gauge bosons become more massive, this contribution is further suppressed due to the decay of the gauge boson wavefunctions. We conclude that  backreaction effects from physical gauge boson production are unimportant in comparison to the effects studied here. \footnote{In the case of the standard model, this backreaction makes the Higgs a little less unstable during high-scale inflation. However, it is numerically subdominant to the backreaction from top quarks, computed in Ref.~\cite{Rodriguez-Roman:2018swn}.}

%%%%%%%%%%%%%%%%%%%%
\section{Standard Model Higgs During Inflation}
\label{sec:inflation}
%%%%%%%%%%%%%%%%%%%%

In this section, we examine the consequences of the enlarged field space for the evolution of the probability distribution governing the SM Higgs vev during inflation, contrasting the case of a scalar field with four real degrees of freedom with the $N=1$ description used in recent studies of the Higgs instability scale, Refs.~\cite{Espinosa:2015qea,Kearney:2015vba,Kohri:2017iyl,East:2016anr}.
We find that, in general, properly accounting for the four degrees of freedom in the Higgs multiplet strengthens constraints on the Hubble scale during inflation at the percent level.
%, resulting from the more rapid initial spread of the probability distribution in the enlarged field space.  
Meanwhile, the mean-squared Higgs vev and resulting energy density are substantially enhanced relative to single-field estimates.

Our aim in the present work is to understand how the four degrees of freedom possessed by the Standard Model Higgs boson affect its evolution during inflation.  To isolate the impact of the Higgs's multiple degrees of freedom,  we  compare our calculations to the one-dimensional calculation of Ref.~\cite{East:2016anr}.  Thus we use the approximate Higgs potential
\begin{align}
V(h) = - b_0 \ln \left( \frac{H^2 + h^2}{\sqrt{e} \Lambda_{\mathrm{max}}^2} \right) \frac{h^4}{4},
\label{eq:SM_approx_V}
\end{align}
where $h$ is the Higgs vev, $b_0 = 0.12 / (4 \pi)^2$, and $\Lambda_\mathrm{max} = 3.0 \times 10^{11}$~GeV.  Because part of the unstable regime of the potential may be stabilized by thermal corrections during reheating, we follow Ref.~\cite{East:2016anr} in  considering the probability in the tail beyond the field value $h_{\mathrm{cr}}$, defined through
\begin{align}
h_{\mathrm{cr}} = - \frac{V^\prime(h_{\mathrm{cr}})}{3 H^2}.
\label{eq:h_cr}
\end{align}
Beyond this field value, the slow roll approximation breaks down and fluctuations rapidly fall into the true vacuum.

As the SM Higgs evolves to large vevs, the mass of the electroweak gauge bosons increases and thus so does the mass gap in the physical spectrum.  For the SM Higgs, the mass gap surpasses the Hubble scale for vevs in the vicinity of $h = 2 H/g$, where $g$ is the coupling constant of $\mathrm{SU}(2)_{\mathrm{L}}$.  At this scale, we expect the effective dimensionality of the Higgs field to transition from four to one.  
 As our experience with the quartic potential in Section~\ref{sec:quartic} suggests (and as we  show explicitly for the SM Higgs below), changing the dimensionality of the Higgs field leads to only minor shifts  in the numerical constraint on $H/\Lambda_{\mathrm{max}}$ compared to the one-dimensional calculation. We thus expect that physically relevant values of $H$ must be substantially smaller than the instability scale, $H/\Lambda_{\mathrm{max}}\lesssim 0.1$, in which case $H < g (\Lambda_{\mathrm{max}}) \Lambda_{\mathrm{max}}/2$.  Thus when the SM Higgs reaches the instability scale and the unstable region of the potential beyond, the radial mode is the only remaining light degree of freedom.  Nonetheless, we demonstrate in this section that accounting for the four degrees of freedom in the SM Higgs at small values of $h$ results in percent-level corrections to constraints on the scale of inflation $H$, coming from the more rapid initial spread of the probability distribution in the enlarged field space.
 
 % the SM Higgs to remain fully four-dimensional throughout its evolution leads to only minor shifts  in the numerical constraint on $H/\Lambda_{\mathrm{max}}$ compared to the one-dimensional calculation.  We thus expect that physically relevant values of $H$ must satisfy $H/\Lambda_{\mathrm{max}}\lesssim 0.1$, in which case $H < g (\Lambda_{\mathrm{max}}) \Lambda_{\mathrm{max}}/2$.  

%%%%%%%%%%%%%%%%%%%%%%%%%%%%
\begin{figure}
\includegraphics[width=\columnwidth]{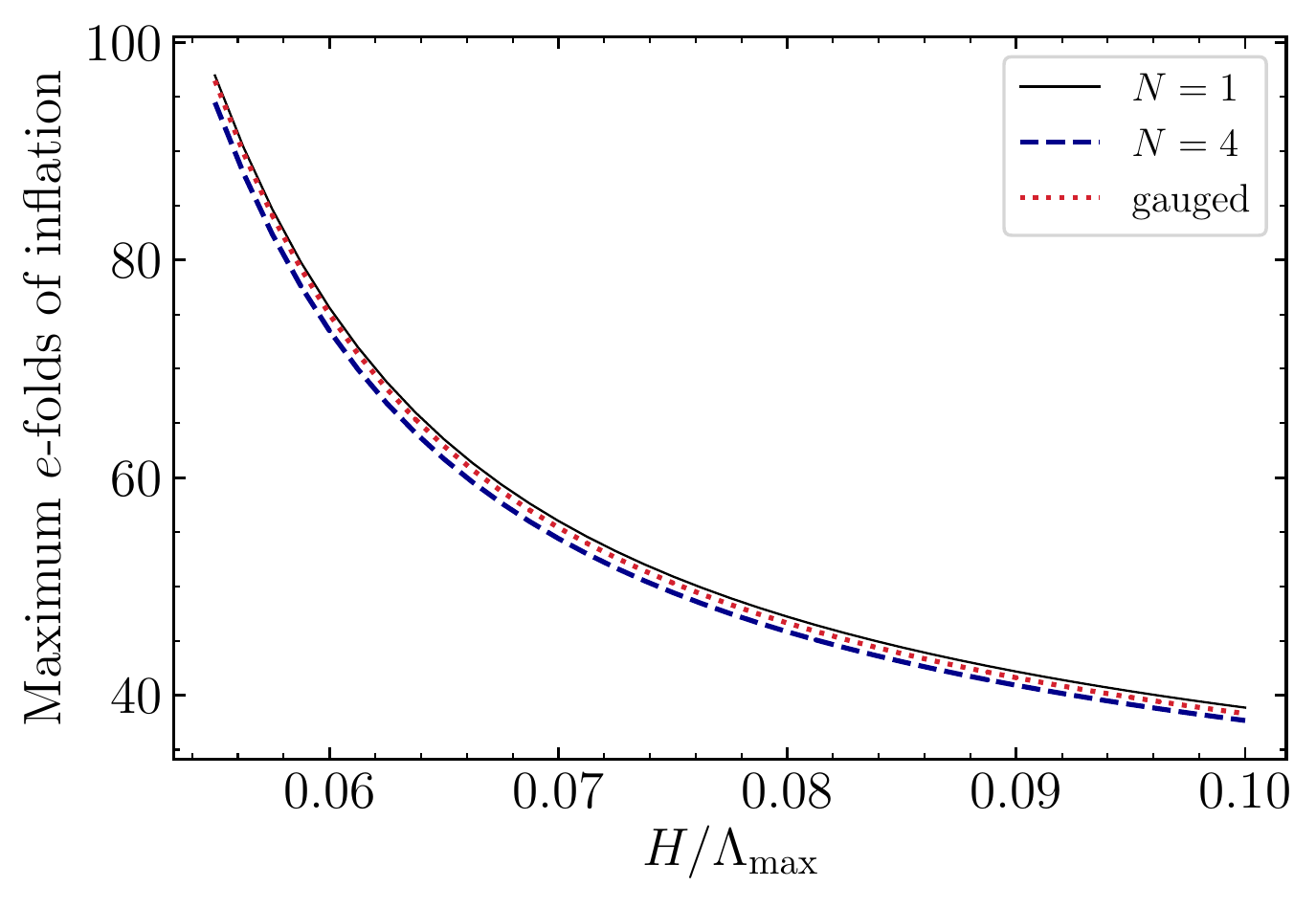}
\caption{The maximum number of $e$-foldings permitted during inflation, as a function of the Hubble parameter during inflation $H$.  Results are shown for scalar fields with $N=1$ (thin black) and $N=4$ (dashed blue) degrees of freedom moving in the SM Higgs potential of Eq.~\eqref{eq:SM_approx_V}.  The dotted red curve (``gauged'') corresponds to a field that interpolates between $N=4$ and $N=1$ degrees of freedom at $2H/g$.
}
\label{fig:SM_Higgs}
\end{figure}
%%%%%%%%%%%%%%%%%%%%%%%%%

To numerically study the evolution of the SM Higgs vev during inflation, we implement a simple model for the transition between $N = 4$ and $N = 1$-dimensional field spaces using a hyperbolic tangent step. Explicitly, we take the effective unitary gauge potential for the radial mode to be
\begin{align}
V(h) = & - b_0 \ln \left( \frac{H^2 + h^2}{\sqrt{e} \Lambda_{\mathrm{max}}^2} \right) \frac{h^4}{4} 
\label{eqn:unitary-gauge-higgs-potential} \\
\nn & - (N-1) \frac{3 H^4}{16 \pi^2} \ln \left(h \right)\(1+\tanh\[\frac{h-2bH/g}{c}\]\)
\end{align}
where $c$ parameterizes the width of the step, and $b$ parameterizes its position, which occurs at $h = 2bH/g$. Unless otherwise noted, we set $c = 8$ and $b=1$; we discuss the dependence of our numerical results on the details of our interpolation below.  In addition, we use a one-loop running gauge coupling $g(h)$~\cite{Martin:1997ns}. 

Details of the numerical calculation are given in Appendix~\ref{sec:comp}. We again consider the evolution of the logarithm of the probability density to ensure accurate results in the tail of the distribution at large $h$, and take the field to intially be a narrowly peaked Gaussian centered at the origin.  The maximum number of $e$-foldings permitted as a function of $H/\Lambda_{\mathrm{max}}$ is shown in Fig.~\ref{fig:SM_Higgs}.  Here we show results for fields with  dimensionality  $N=1$ and $N=4$ held artificially constant throughout the field evolution, as well as the more physical case of a field that interpolates between $N=4$ degrees of freedom for $h/H = \hat \chi \ll 2/g$ and $N=1$ degrees of freedom for $\hat \chi \gg 2/g$ via Eq.~\eqref{eqn:unitary-gauge-higgs-potential}.
Our  results for $N=1$ agree well with those  of Ref.~\cite{East:2016anr}.  For $N=4$, the number of $e$-foldings permitted decreases, corresponding to stricter constraints on $H/\Lambda_{\mathrm{max}}$.  This faster increase of the probability to find the Higgs at catastrophically large vevs for $N=4$ is a consequence of both enhanced probability to find the field at larger vevs as well as more rapid evolution to the asymptotic value.  At small values of $H/\Lambda_{\mathrm{max}}$, where the probability distribution function is able to evolve for more $e$-foldings, the difference between $N=1$ and $N=4$ is more pronounced, as expected for the faster evolution seen in larger dimensional spaces.  Overall, however, the differences between $N=1$, $N=4$, and the physical unitary gauge scenario are numerically small.

Observations typically require around 60 $e$-folds of inflation to solve the horizon problem~\cite{Guth:1980zm}, which for $N=1$ yields the condition $H / \Lambda_\mathrm{max} \lesssim 0.067$, in agreement with Ref.~\cite{East:2016anr}.  However, when we account for the full degree of freedom content of the Standard Model Higgs, the constraint is strengthened slightly: our unitary gauge calculation requires $H / \Lambda_\mathrm{max} \lesssim 0.066$.  Consideration of this effect is thus important for precision determination of constraints on the inflationary scale $H$.  We have checked that the unitary gauge results in Fig.~\ref{fig:SM_Higgs} are insensitive to the details of the transition from $N = 4$ to $N = 1$, and in particular are insensitive to the specific values adopted for the parameters $b$ and $c$ in Eq.~\eqref{eqn:unitary-gauge-higgs-potential}.
The results of Fig.~\ref{fig:SM_Higgs}  do depend weakly on the choice of initial condition.  As the initial condition becomes increasingly peaked at the origin, the number of allowed $e$-folds asymptotes to a fixed value, and thus for sufficiently narrow Gaussian initial conditions, the resulting constraint on the Hubble scale becomes independent of the initial conditions. We have checked that the initial conditions used in Fig.~\ref{fig:SM_Higgs}, given in Eqs.~\eqref{eqn:init-condn-form} and~\eqref{eqn:initial-condition-width}, yield constraints on the number of allowed $e$-folds that agree with the asymptotic values at the sub-percent level.  
 For insufficiently peaked Gaussian initial conditions, however, the field expands more rapidly toward the unstable region of the potential, and accordingly fewer $e$-folds are allowed. 

%%%%%%%%%%%%%%%%%%%%%%%%%%%%%
\begin{figure}[t!]
\includegraphics[width=\columnwidth]{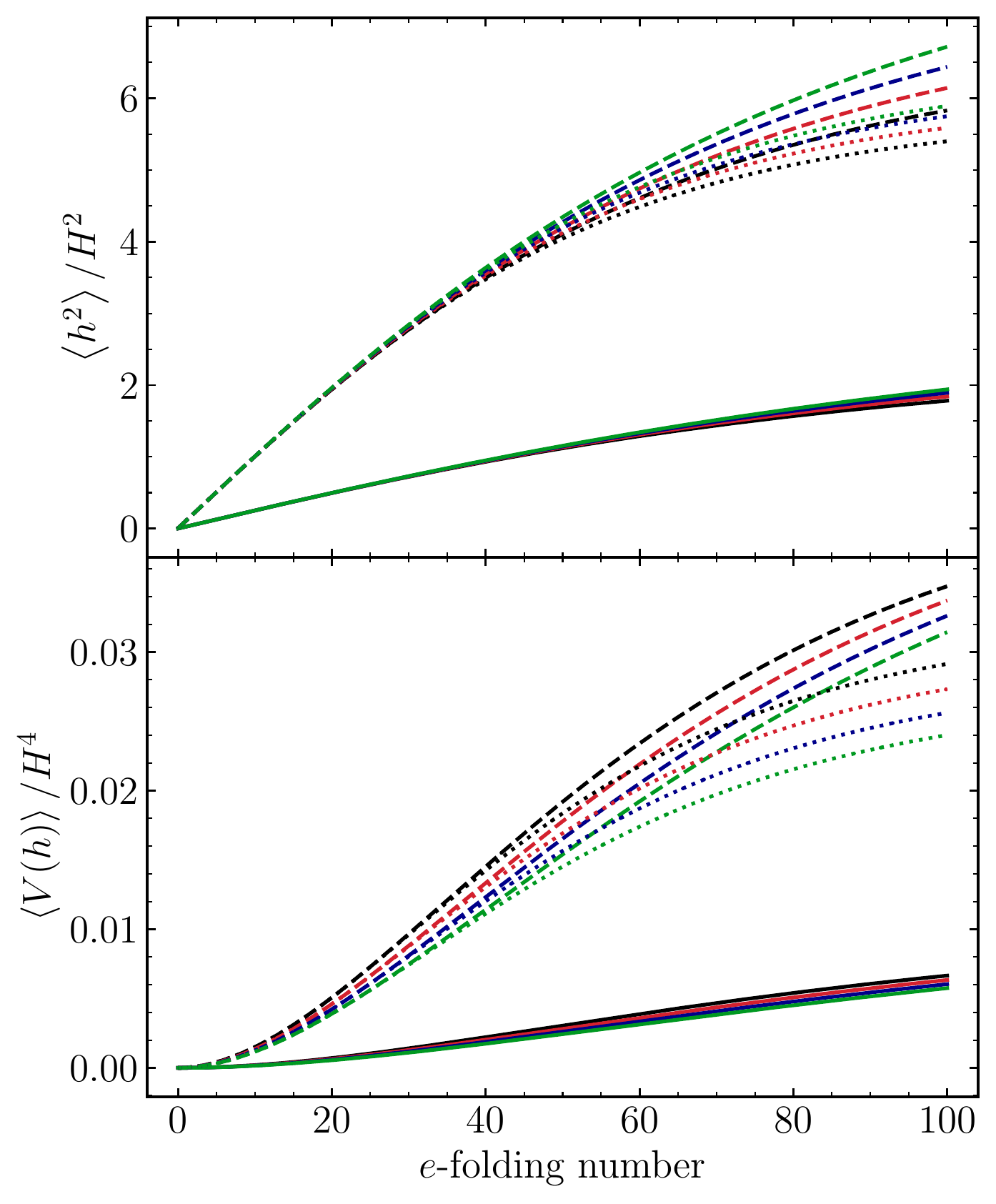}
\caption{
The average vev-squared of the Higgs field (top) and energy density in the Higgs field (bottom) as a function of $e$-folding number for a one-dimensional field (solid lines), four-dimensional field (dashed lines), and a unitary-gauge field  (dotted lines) evolving in a radial potential given by Eq.~\eqref{eqn:unitary-gauge-higgs-potential}. We show the result for $H/\Lambda_{\rm max}$ taking values $0.055$ (black), $0.07$ (red), $0.085$ (blue), and $0.1$ (green) (from top to bottom for each set of curves on the upper panel, and from bottom to top on the lower panel). For these results, we set the interpolation parameters $b = 1$ and $c = 8$.
}
\label{fig:SM_Higgs_energy}
\end{figure}
%%%%%%%%%%%%%%%%%%%%%%%%%%%%%

Although  the enlarged dimensionality of the physical SM Higgs field space has only a small effect on the allowed scale of inflation, it has a more dramatic effect on the vev and the energy density stored in the field, similar to the results for the quartic potential in Sec.~\ref{sec:quartic}. In Fig.~\ref{fig:SM_Higgs_energy} we show the evolution of the vev and  potential energy in the Higgs field computed from Eq.~\eqref{eq:h_cr}.
The four-dimensional field shows substantially larger vevs and energy densities than the one-dimensional field. Increasing $N$ from 1 to 4 increases the energy density by a factor of $\sim 4$, and the vev-squared by $\sim 3$, broadly in line with the scalings derived for the quartic potential in Sec.~\ref{sec:quartic}.  However, a four-dimensional field moving in the SM Higgs potential also displays a more substantial dependence on the value of $H/\Lambda_{\rm max}$, reflecting the more rapid evolution of the tails of the probability distribution function toward the non-quartic regions of the potential.   The running of $\lambda$ with the vev means that, for larger $H / \Lambda_\mathrm{max}$, the field experiences a flatter potential at large $\hat{\chi} = \chi/H$.  Thus for larger $H / \Lambda_\mathrm{max}$,
the Higgs is able to wander out to larger vevs relative to Hubble, while the expected energy density (in units of  $H^4$) in the Higgs field decreases with increasing $H/\Lambda_\mathrm{max}$.

\begin{figure*}[t!]
	\includegraphics[width=\textwidth]{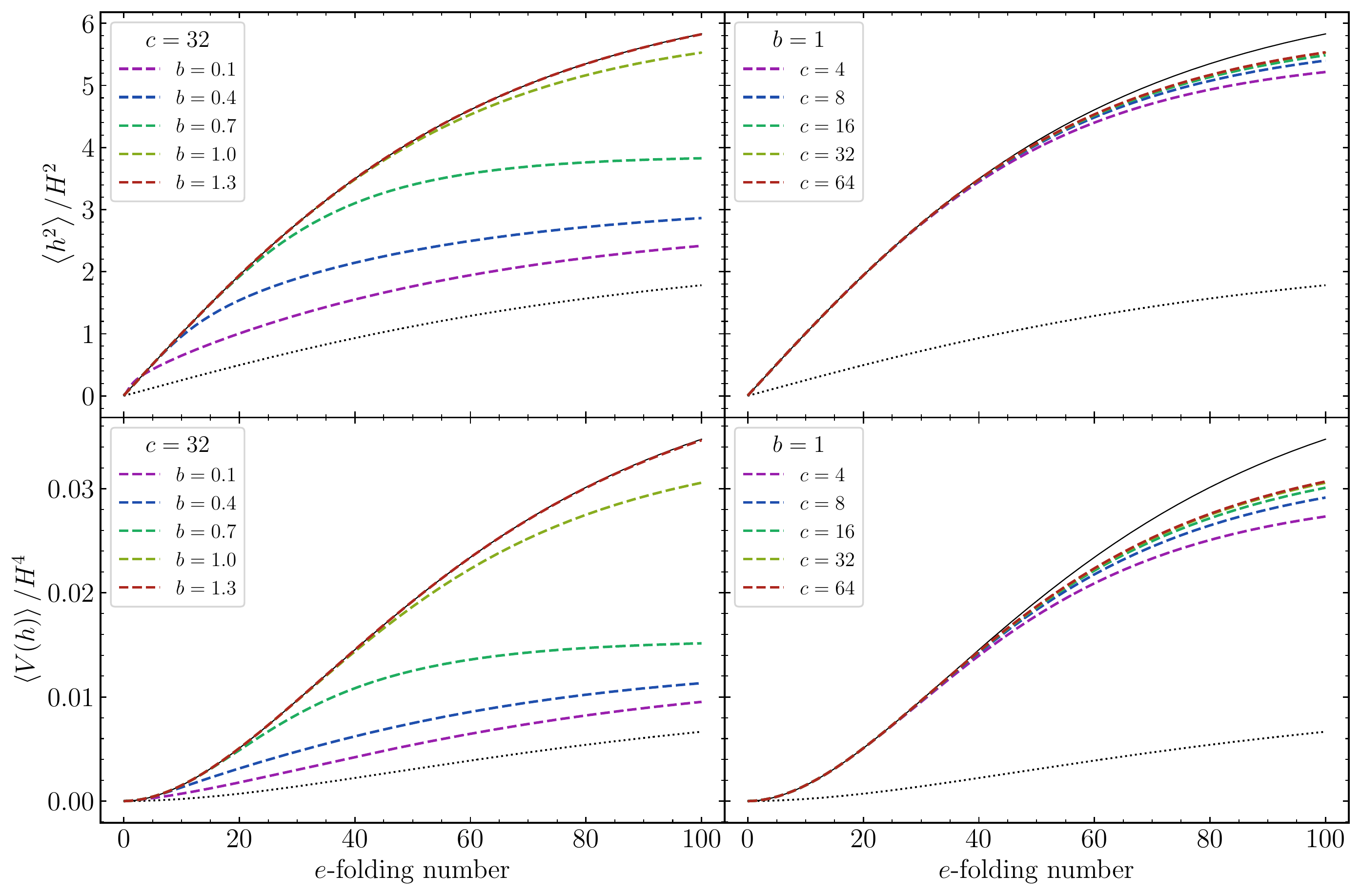}
	\caption{
		The evolution of the squared vev (top panels) and the averaged energy density (bottom panels) for fixed $H/\Lambda_{\rm max} = 0.055$,
		varying the location and steepness of the transition function.
		In all cases, the lower, dotted and upper, solid black curves denote the ungauged $N=1$ and $N=4$ results, respectively.
		For the dashed colored curves, the left panels fix the steepness $c = 32$ and vary the location $b$ from $0.1$ (lower, violet) to $1.3$ (upper, red) in steps of $\Delta b = 0.3$.
		The right panels fix $b = 1$ and vary $c$ from $c = 4$ (lower, green) through $c = 64$ (upper, red) in factors of 2.
	}
	\label{fig:SM_Higgs_energy_transition}
\end{figure*}

As Fig.~\ref{fig:SM_Higgs_energy} shows, the unitary gauge field (with our default implementation of the transition, Eq.~\eqref{eqn:unitary-gauge-higgs-potential}) realizes a vev and an energy density much closer to those of the pure four-dimensional field than the one-dimensional field. Unfortunately,  results for the vev and energy density are somewhat sensitive to the details of how the interpolation from four- to one-dimension in Eq.~\eqref{eqn:unitary-gauge-higgs-potential} is implemented, as we show in more detail in Fig.~\ref{fig:SM_Higgs_energy_transition}. In particular, our results for the vev and energy density are relatively insensitive to the steepness of the transition, but they depend in detail on its location.  This dependence occurs because the SM gauge bosons become massive before the Fokker-Planck equation is dominated by the gradient flow of the SM Higgs potential (see, e.g.\ Eq.\ \eqref{eqn:FPeqnspherical}). Moving the transition thus directly affects the duration of the evolution in the four-dimensional potential.   Going beyond the parameterization of Eq.~\eqref{eqn:unitary-gauge-higgs-potential} to a more precise calculation of the form of the turnover requires explicit loop computations in de Sitter space and is beyond the scope of this work; regardless of the remaining uncertainties, however, Fig.~\ref{fig:SM_Higgs_energy}  makes it clear that it is critical to include the full dimensionality of the Higgs field to estimate the mean properties of the field during inflation.

%%%%%%%%%%%%%%%%%%%%%%%%%%%
\section{Conclusion}
\label{sec:conclusion}
%%%%%%%%%%%%%%%%%%%%%%%%%%%

In this work, we have studied the stochastic evolution of the vev of a scalar field transforming in a linear representation of a continuous symmetry group in a de Sitter background. We generalized the derivation of the single-field Fokker-Planck equation to scenarios where the vev takes values in a flat, multi-dimensional field space. For gauged symmetries, we demonstrated how this multi-dimensional evolution is recovered in unitary gauge.

As a useful example, we considered in detail a scalar field with a pure quartic potential, which admits an analytical solution for the asymptotic probability distribution. Enlarging the dimensionality $N$ of the field space has the important effects of (i) increasing the vev (as $\sim N^{1/4}$), (ii) increasing the energy density stored in the scalar field (as $ N$), and (iii) making the approach to the asymptotic probability distribution more rapid.

Applying these insights to the more subtle case of the SM Higgs, we found that accounting for the enlarged field space of the physical Higgs boson near the origin in field space strengthens constraints on the scale of inflation at the percent level.  Although the effect on the probability of the Higgs to obtain catastrophically large field values is small, the enlarged field space has a significant effect on the vev and energy density.  Larger predicted vevs may have a significant effect on the duration of reheating~\cite{Freese:2017ace}, with potential observational consequences in non-gaussianities created during Higgs-modulated reheating~\cite{Lu:2019tjj}, as well as on models which rely on the inflationary Higgs vev for baryogenesis~\cite{Kusenko:2014lra,Pearce:2015nga,Yang:2015ida}.

%\textcolor{red}{While in this work we have considered only the Standard Model Higgs boson in purely anti-de Sitter space, it has been observed that departures from pure anti-de Sitter space and/or higher dimensional effective operators in the Higgs potential can also modify bounds on Standard Model parameters at the percent level~\cite{Fumagalli:2019ohr}.  We leave the incorporation of the effects discussed here into these models for later work.}

The techniques developed here are applicable to any light scalar field, including curvatons~\cite{Linde:1996gt,Enqvist:2001zp,Lyth:2001nq,Moroi:2001ct}.  The enhanced vev, and particularly its faster approach to its equilibrium value, may be particularly relevant to these scenarios.  Such a curvaton may evade the concerns of Ref.~\cite{Hardwick:2017fjo}, which demonstrated that the probability density of a real scalar curvaton may depart significantly from its de Sitter equilibrium form during slow roll inflation.  We leave the full study of the vev of a multi-dimensional curvaton and its fluctuations for future work.   Other interesting avenues for future investigation include studying the impact of the evolving dimensionality of the SM Higgs field on inflationary constraints in the presence of higher-dimensional operators, particularly since higher-dimensional operators have been shown to magnify the impact of small deviations from the benchmark stochastic Higgs analysis in some cases \cite{Fumagalli:2019ohr}.

\acknowledgments

We thank Patrick Draper, Mark Hertzberg, and Lian-Tao Wang for useful discussions. The work of P.A.\ and L.P.\ was supported by the US Department of Energy through grant DE-SC0015655. The work of J.S.\ was supported in part by DOE Early Career grant DE-SC0017840.
Z.J.W.\ is supported in part by the United States Department of Energy Computational Science Graduate Fellowship, provided under Grant No. DE-FG02-97ER25308.
L.P.\ and J.S.\ thank the Aspen Center for Physics for hospitality and support
through National Science Foundation grant PHY-1607611. P.A.\ thanks the  Yukawa Institute for Theoretical Physics at Kyoto University, where some of this work was completed during the YITP-T-19-02 on ``Resonant instabilities in cosmology.''
P.A.\ and Z.J.W.\ acknowledge the hospitality of the Kavli Institute for Theoretical Physics, which is supported in part by the National Science Foundation under Grant No.\ NSF-PHY-1748958.

%%%%%%%%%%%%%%%%%%%%%%%%%%%%%%%%%%%%%%%%
\appendix

%%%%%%%%%%%%%%%%%%%%%%%%%%%
\section{Eigenfunctions and eigenvalues for excited modes in a quartic potential}
\label{sec:lneq0}
%%%%%%%%%%%%%%%%%%%%%%%%%%%

In this Appendix, we study the excited eigenfunctions and eigenvalues in the quartic potential. We begin by writing down the general Fokker-Planck equation for an SO(N) invariant potential, before showing numerically that higher dimensional field spaces lead to larger eigenvalues.

As Eq.~\eqref{eq:rho1} shows, the contribution of each eigenmode to the probability distribution decays away as $e^{-\Lambda_n (t-t_0)}$.  Consequently, the asymptotic distribution is given by the eigenmode with the eigenvalue of zero, while the approach to equilibrium is determined by the remaining eigenmodes.  As we are interested in describing the approach to equilibrium, we consider in this appendix the higher eigenmodes, characterized by non-zero eigenvalues, $\Lambda_{n,m,\dots} \neq 0$, for a quartic potential.

We separate the mode functions $\Phi_{n,m,\dots}$ into radial and angular parts.  It is convenient to consider separately $N=2$, which has only an azimuthal angle, and $N\geq 3$, which additionally has polar angle(s), although for rotationally symmetric initial conditions, only rotationally invariant modes contribute to $\rho(\chi_L)$.

As in the case of a real scalar field, it is convenient when considering a quartic potential to  introduce the rescaled variables~\cite{Starobinsky:1994bd}
\begin{align}
\vec{\chi}_L = H \lambda^{-1 / 4} \vec{\tilde{\chi}}, \quad \text{and}\quad \Lambda_{n,\dots} = \lambda^{1 / 2} H \tilde{\Lambda}_{n,\dots}  .
\label{eq:Phi_DE1}
\end{align}
We also introduce the rescaled mode functions
\begin{align}
\tilde \Phi_{n,m,\dots}(\vec{\tilde{\chi}}) \equiv \left( \frac{H}{\lambda^{1 / 4}} \right)^{N / 2} \Phi_{n,m,\dots}(\vec{\tilde{\chi}})
\end{align}
which obey the normalization condition
\begin{align}
\int \ud^N \tilde \chi \, \tilde \Phi_{n,m,\dots}(\vec{\tilde \chi}) \tilde \Phi_{n^\prime,m^\prime,\dots}^\dag(\vec{\tilde \chi}) &= \delta_{n,n^\prime} \delta_{m,m^\prime} \dots.
\end{align}
From Eq.~\eqref{eq:eigenfunction1}, we see that these mode functions satisfy
\begin{align}
- \tilde{\nabla}^2 \tilde \Phi_{n,m,\dots}
+ \left[
(\vec{\tilde{\nabla}} v)^2 - \tilde{\nabla}^2 v
\right]\tilde  \Phi_{n,m,\dots}
&= 8 \pi^2 \tilde{\Lambda}_{n,\dots} \tilde \Phi_{n,m,\dots},
\label{eq:Phi}
\end{align}
where all explicit $\lambda$ dependence has been removed by the rescaling.  A straightforward generalization of the argument in Ref.~\cite{Starobinsky:1994bd} shows that all eigenvalues  $\tilde{\Lambda}_{n,\dots}$ are non-negative.

The probability distribution over the rescaled field is  $\tilde \rho(\tilde{\chi}) = ( H / \lambda^{1 / 4} )^N \rho(\vec \chi_L)$, with normalization condition
\begin{align}
\int \tilde{\rho}(\vec{\tilde{\chi}}) \, \ud^N \tilde{\chi} &= 1 .
\label{eq:norm2}
\end{align}
The expansion coefficients in the rescaled probability distribution are given by $\tilde a_{n,m,\dots} = (H / \lambda^{1 / 4})^{N / 2} a_{n,m,\dots} $, where $ a_{n,m,\dots}$ are the expansion coefficients in Eq.~\eqref{eq:rho1}.

In the following subsections, we first study the eigenvalue equations for $N=2$ and $N \geq 3$.  We then demonstrate in Sec.~\ref{ap:eigenmodes_nonzero_L} that modes with $L \neq 0$ are unimportant if the initial condition is sufficiently localized near the origin.  Finally, we present numerical studies of the lowest three eigenfunctions in Sec.~\ref{ap:eigenmodes_numerical}.

\subsection{$N=2$ $(\mathrm{SO}(2) \cong \mathrm{U}(1))$}
\label{ap:eigenmodes_N2}

We now consider the excited eigenmodes of a scalar field with two real degrees of freedom, which is appropriate to a field with a $\mathrm{SO}(N)$ or $\mathrm{U}(1)$ global symmetry.  Again, we specialize  to the quartic potential.  We use polar coordinates on field space, with radial coordinate $\tilde{\chi} = \sqrt{ \vec{\tilde{\chi}} \cdot \vec{\tilde{\chi}}}$ and angular coordinate $\phi$.  We introduce the ansatz
\begin{align}
\tilde \Phi_{n,\ell}(\vec{\tilde{\chi}}) &= R_{n,\ell}(\tilde{\chi}) Y_\ell(\phi),
\end{align}
where as usual
\begin{align}
Y_\ell(\phi) &= \frac{1}{\sqrt{2\pi}} e^{i \ell \phi},
\end{align}
with $\int_0^{2\pi} \ud \phi \, Y_\ell(\phi) Y_{\ell^\prime}^*(\phi) = \delta_{\ell,\ell^\prime}$.

Due to the rotational symmetry of the Lagrangian, the potential, and hence $v$, depends only on $\tilde{\chi}$.  Therefore, $R_{n,\ell}$ satisfies the equation
\begin{align}
8 \pi^2 \tilde{\Lambda}_{n\ell} R_{n,\ell} &= - \left[ R_{n,\ell}^{\prime \prime}
+ \frac{R_{n,\ell}^\prime}{\tilde{\chi}}
- \frac{\ell^2}{\tilde{\chi}^{2}} R_{n,\ell}
\right] \nonumber \\
& \qquad + \left[ (v^\prime)^2 - v^{\prime \prime} - \frac{v^\prime}{\tilde{\chi}} \right] R_{n,\ell} .
\label{eq:N2}
\end{align}
For the quartic potential, this is
\begin{align}
8 \pi^2 \tilde{\Lambda}_{n\ell} R_{n,\ell} &=
-R_{n,\ell}^{\prime \prime}
- \frac{R_{n,\ell}^\prime}{\tilde{\chi}}
+ \frac{\ell^2}{\tilde{\chi}^{2}} R_{n,\ell} \\ \nonumber
& \qquad
+ \left[ \frac{16 \pi^4 \tilde{\chi}^6}{9}
- 4 \pi^2 \tilde{\chi}^2
- \frac{4 \pi^2 \tilde{\chi}^2}{3}
\right] R_{n,\ell} .
\end{align}
This equation can be solved numerically to find the eigenvalues and their corresponding eigenmodes, as we discuss further below.

\subsection{$\mathrm{SO}(N)$ with $N\geq 3$}
\label{ap:eigenmodes_N3}

We now turn our attention to scalar fields with more than three real degrees of freedom.  Our coordinates are $\tilde{\chi} = \sqrt{ \vec{\tilde{\chi}} \cdot \vec{\tilde{\chi}}}$, which serves as a radial coordinate in the field space, and the angular variables $\phi,\theta_1,\theta_2,\dots$, which characterize positions on the sphere $S^{N-1}$.  The Laplacian operator can be decomposed as:
\begin{align}
\tilde{\nabla}^2
&= \frac{\partial^2}{\partial \tilde{\chi}^2} + \frac{N-1}{\tilde{\chi}} \frac{\partial}{\partial \tilde{\chi}} + \frac{1}{\tilde{\chi}^{2}} \nabla^2_{S^{N-1}} ,
\end{align}
where $\nabla^2_{S^{N-1}}$ is the Laplacian on the $N-1$-sphere.  This operator can be expressed inductively, starting with the familiar $2$-sphere case,
\begin{align}
\nabla_{S^2}^2  &= \frac{1}{\sin\theta_1} \frac{\partial}{\partial \theta_1} \left( \sin\theta_1 \frac{\partial }{\partial \theta_1} \right) + \frac{1}{\sin^2\theta_1} \frac{\partial^2 }{\partial \phi^2},\\ \nonumber
\nabla_{S^{N+1}}^2  &= \frac{1}{\sin^N\theta_N} \frac{\partial}{\partial \theta_N} \left( \sin^N\theta_N \frac{\partial }{\partial \theta_N} \right) + \frac{1}{\sin^2\theta_N} \nabla^2_{S^{N}} .
\end{align}
The eigenfunctions of $\nabla^2_{S^{N-1}}$ are generalized (or scalar) spherical harmonics, which have the eigenvalues
\begin{align}
\nabla^2_{S^{N-1}} Y_{m,\ell_1,\ell_2,\dots}(\phi,\theta_1,\theta_2,\dots) = - L (L + N - 2).
\end{align}
where $L = \ell_{\mathrm{max}}$ is the maximum of $\ell_1,\ell_2,\dots$.  These spherical harmonics can be expressed in terms of Legendre functions~\cite{Higuchi:1986wu},
\begin{align}
Y_{m,\ell_1 \leq \ell_2 \leq \dots L}(\phi, \theta_1, \theta_2, \dots)
&= \frac{1}{\sqrt{2\pi}} e^{i m \phi}
\prod_{j=1}^{N-2} \phantom{}_{j+1} \bar{P}_{\ell_j}^{\ell_{j-1}} (\theta_j),
\end{align}
where $\ell_0 \equiv m$, $|m| \leq \mathrm{min}(\ell_i)$ and
\begin{align}
\phantom{}_j \bar{P}^\ell_L(\theta)
&= \sqrt{ \frac{2L + j -1}{2} \frac{(L + \ell + j -2)!}{(L-\ell)!}}
(\sin\theta )^{\frac{(2-j)}{ 2}}\nonumber \\
& \qquad \times  P_{L + \frac{j-2}{2}}^{-\left( \ell + \frac{j-2}{2} \right)} (\cos\theta).
\end{align}
These spherical harmonics  satisfy the normalization condition\footnote{These agree with the familiar $N=3$ spherical harmonics up to factors of $(-1)^m$, which are the Condon-Shortly phases and do not affect orthonormality.}
\begin{align}
\int \ud\Omega^{N-1} \, Y_{m,\ell_1,\ell_2,\dots}
Y^*_{m^\prime,\ell_1^\prime,\ell_2^\prime,\dots}
  &= \delta_{m,m^\prime} \delta_{\ell_1,\ell_1^\prime}\dots.
\end{align}

We make the ansatz that the mode functions may be decomposed in terms of these harmonics as
\begin{align}
\tilde \Phi_{n,m,\dots}(\tilde{\chi}, \phi, \theta_1, \dots) = R_{n,L}(\tilde{\chi}) Y_{m,\ell_1,\dots}(\phi,\theta_1,\dots).
\end{align}
Using the rotational invariance of $v$, $R_{n,L}$ satisfies the differential equation
\begin{align}
& 8 \pi^2 \tilde{\Lambda}_{n,L} R_{n,L} =
\left[ (v^\prime)^2 - \left( v^{\prime \prime} + \frac{N-1}{\tilde{\chi}} v^\prime \right) \right] R_{n,L}  \\ \nn
& \quad  - \left[ R_{n,L}'' + \frac{N-1}{\tilde{\chi}} R_{n,L}' - \frac{L}{\tilde{\chi}^2}
\left( L + N - 2 \right) R_{n,L} \right] .
\label{eq:R_eqn}
\end{align}
Note that setting  $N=2$ and $L = \ell$ recovers Eq.~\eqref{eq:N2}.

Again this differential equation can be studied numerically, which we discuss below.

\subsection{Contribution of $L \neq 0$ to the probability density}
\label{ap:eigenmodes_nonzero_L}

Modes with $L \neq 0$ do not contribute if the initial conditions are spherically symmetric.  Recall that the contribution from each mode is determined by the coefficients $\tilde a_{n,m,\dots}$.  In general, the coefficients are determined by the initial condition at time $t = t_0$, as
\begin{align}
\tilde a_{n,m,\dots} &= \int \ud^N \tilde \chi \, \tilde \rho(\vec{\tilde \chi},t_0) e^{v(\vec{\tilde \chi})} \tilde \Phi_{n,m,\dots}(\vec{\tilde \chi}).
\end{align}
The orthogonality of the generalized spherical harmonics immediately imposes that $\tilde a_{n,m,\dots}$ is zero for $L \neq 0$ if the initial state is spherically symmetric.

We can show more generally that if the initial vev is localized near the origin, the $L=0$ contribution dominates the probability density $\rho$.  Since the vev is localized, we seek the behavior of $R_{n,L}(\tilde{\chi})$ in the small $\tilde{\chi}$ limit. In this limit, Eq.~\eqref{eq:R_eqn} becomes
\begin{align}
- R''_{n,L} - \frac{N-1}{\tilde{\chi}} R_{n,L}^\prime
+ \frac{L (L+N-2)}{\tilde{\chi}^2} R_{n,L}
 &= 0.
\end{align}
As can be seen from Eq.~\eqref{eq:N2}, this equation also holds for $N=2$ with $L=1$.  Because $L + N - 2 \geq 0$, the differential equation has the solution
\begin{align}
R_{n,L} &=
c_1 \tilde \chi ^L
+ c_2 \tilde \chi ^{-L-N+2}.
\end{align}
For $L > 0$, we must have $c_2 = 0$, because $\tilde  \chi ^{-L-N+2} \rightarrow \infty$ as $\chi \to 0$, in which case $c_2\neq 0$ would make the eigenfunction non-normalizable.  This leaves us with
\begin{align}
R_{n,L} \propto \tilde \chi^L
\end{align}
at small $\tilde \chi$.  Since $\tilde{a}_{n,m,\ell_1,\dots}$ is determined by the overlap of the mode with the initial condition, it is suppressed for $L \neq 0$ when the initial condition is localized near the origin.  Consequently, we ignore eigenmodes with $L \neq 0$ throughout this work.

We note that this follows from Eq.~\eqref{eq:R_eqn} and does not depend on the functional form of $v$.  Therefore, this result applies to any potential that depends only on the radial field coordinate.

\begin{figure}[th]
\begin{center}
\includegraphics[width=\columnwidth]{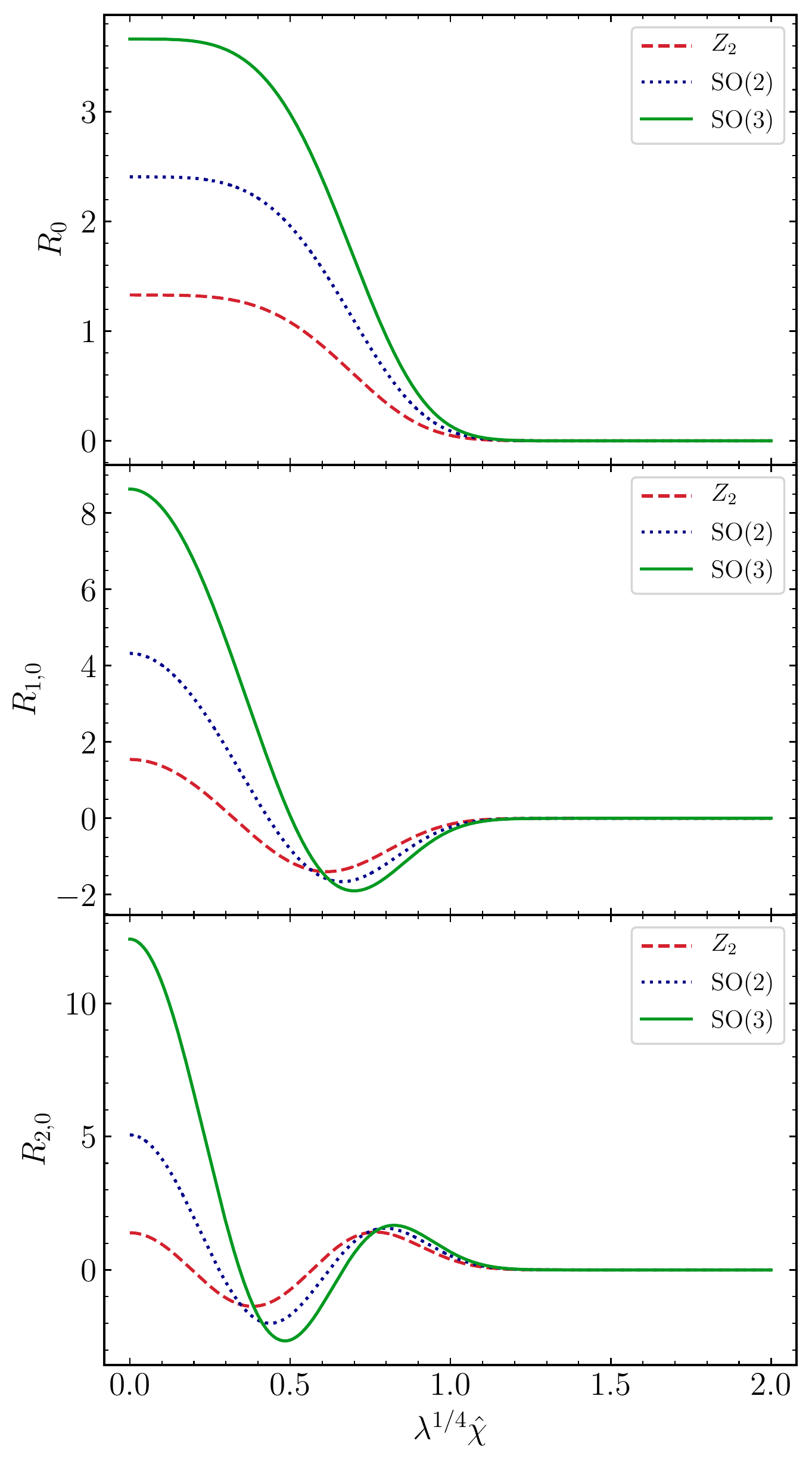}
\caption{$R_{0}(\tilde{\chi})$ (top), $R_{1,0}(\tilde{\chi})$ (middle), $R_{2,0}(\tilde{\chi})$ (bottom), for a real field with $Z_2$ symmetry, $\mathrm{SO}(2)$ symmetry, and $\mathrm{SO}(3)$ symmetry. }
\label{fig:R_plots}
\end{center}
\end{figure}

\subsection{Numerical Mode Functions }
\label{ap:eigenmodes_numerical}

 In this subsection we present a numerical study of the eigenmodes $R_{n,0}$ for $n > 0$.  For numerical computation, it is convenient to work with the functions
\begin{align}
R_{n,L}(\tilde{\chi}) = G_{n,L}(\tilde{\chi}) e^{-v(\tilde{\chi})} ,
\end{align}
which satisfy the differential equation
\begin{align}
8 \pi^2 \tilde{\Lambda}_{n,L}  G_{n,L} &= - G_{n,L}^{\prime \prime} + 2 v^\prime G_{n,L}^\prime
- \frac{N-1}{\tilde{\chi}} G_{n,L}^\prime  \nonumber \\
& \qquad
+ \frac{L(L+N-2)}{\tilde{\chi}^2} G_{n,L}.
\label{eq:G_n_DE}
\end{align}

For a single field, the first two non-zero eigenvalues are $\tilde{\Lambda}_{1,0} = 0.0889$ and $\tilde{\Lambda}_{2,0} = 0.289$ (in agreement with Ref.~\cite{Starobinsky:1994bd}).  For $\mathrm{SO}(2)$, we find that the lowest eigenvalues are $\tilde \Lambda_{1,0} = 0.3656$, and $\tilde \Lambda_{2,0} = 0.933$, and for $\mathrm{SO}(3)$, we find $\tilde{\Lambda}_{1,0} = 0.4344$, and $ \tilde{\Lambda}_{2,0} = 1.034$.  As $N$ increases, so do the corresponding eigenvalues.  Since the eigenvalues enter the probability distribution function $\rho$ through the factor $e^{- \Lambda_n (t- t_0)}$, this increase in the eigenvalues causes the contribution of the higher modes to decay away more rapidly with increasing $N$.

Finally, for completeness, we discuss the corresponding eigenfunctions, shown in Fig.~\ref{fig:R_plots}.  As the symmetry group becomes larger, features in the mode functions are shifted to larger $\tilde{\chi}$ and the vev accordingly wanders to larger values, as we saw for the zero mode.

%%%%%%%%%%%%%%%%%%%%%%%%%%%%
\section{Numerical methods}
\label{sec:comp}
%%%%%%%%%%%%%%%%%%%%%%%%%%%%

%%%%%%%%%%%%%%%%%%%%%%%%%%%%%

In this appendix we present our numerical implementation to solve the Fokker-Planck equation.
The scalar potential for fixed one- and four-dimensional theories is given in Eq.~\eqref{eq:SM_approx_V}, and for unitary gauge in Eq.~\eqref{eqn:unitary-gauge-higgs-potential}, and the field value which solves Eq.~\eqref{eq:h_cr} defines the tail of the distribution.
We also solve the Fokker-Planck equation for the pure quartic potential, Eq.~\eqref{eqn:quarticpot}.
In each case, we numerically solve the differential equation for the natural logarithm of the probability distribution,\footnote{
    When studying the SM Higgs in unitary gauge, we rescale the probability distribution by a factor of $\chi^3$ to match that for the $N=4$ case, as (initially) $\rho$ has most of its support in the region that is effectively 4-dimensional.
    Note that this rescaling also alters the boundary condition specified below.
}
\begin{align}
\frac{\partial X}{\partial \hat t}
&= \frac{(N+2 \hat \chi  v^\prime-1) }{8 \pi ^2\hat  \chi }
\frac{\partial X}{\partial \hat \chi}
+ \frac{1}{8 \pi^2} \left( \left( \frac{\partial X}{\partial \hat \chi} \right)^2 + \frac{\partial^2 X}{\partial \hat \chi^2} \right) \nonumber \\
& \qquad +\frac{(N-1) v^\prime+\hat \chi v^{\prime \prime}}{4 \pi ^2 \hat \chi },
\label{eqn:fp-for-X}
\end{align}
where $\hat \chi = h / H$ is the rescaled vev.

For our initial condition, we set the probability distribution as an $n$-dimensional Gaussian,
\begin{align}
    \rho(h,0)
    &= \frac{1}{\left( 2 \pi \left\langle h^2 \right\rangle \right)^{N/2}}
        \exp \left(- \frac{h ^2}{2 \left\langle h^2 \right\rangle} \right)
\label{eqn:init-condn-form}
\end{align}
where
\begin{align}\label{eqn:initial-condition-width}
\left< h^2 \right> &= \frac{H^2 \tanh \left(\frac{\sqrt{2 \lambda }}{8 (2 \pi )}\right)}{(2 \pi ) \sqrt{2 \lambda }}.
\end{align}
For the SM Higgs, $\lambda$ can be found by evaluating Eq.~\eqref{eq:SM_approx_V} at $H=0$.
With $N=1$, this is identical to the initial condition used in Ref.~\cite{East:2016anr}.

We numerically solve Eq.~\eqref{eqn:fp-for-X} using \textsf{SciPy}'s DOP853 routine~\cite{scipy,10.5555/153158} with fourth-order finite differencing for derivatives with respect to $\hat{\chi}$.
We impose the same boundary conditions as~\cite{East:2016anr}, namely, that $\partial X / \partial \chi = \chi \partial^2 X / \partial \chi^2$, which is satisfied by the Gaussian initial condition.
Because the initial conditions are sharply peaked near $\chi = 0$, we use a grid spanning $(0, \hat{\chi}_\mathrm{max}]$ whose spacing increases geometrically from $\hat{\chi}_\mathrm{max} / N_1$ at $\hat{\chi} = 0$ to $\hat{\chi}_\mathrm{max} / N_2$ at $\hat{\chi} = \hat{\chi}_\mathrm{max}$, in practice choosing $N_1 = 8 N_2$. Doing so provides an excellent compromise between computational cost and accuracy and produces solutions which satisfy the probability normalization constraint to a higher precision than using a uniform grid.

For the pure-quartic and SM Higgs models we choose $N_2 = 384$ and $512$, respectively, checking that all of our results are consistent with $N_2 = 256$ and $384$.
For the quartic model, we set $\hat{\chi}_\mathrm{max}$ to be a multiple of the critical value predicted analytically by Eq.~\eqref{eq:chicr1}, finding a factor of $1.2$ to be sufficiently large.
Similarly, for the SM Higgs model we set $\hat{\chi}_\mathrm{\max}$ to be $1.4$ times the critical value defined in Eq.~\eqref{eq:h_cr}.
In both cases we verified that increasing the size of the domain has a negligible effect on our results; in particular, the results presented in Figs.~\ref{fig:chi_-180} and~\ref{fig:SM_Higgs} change by no more than one part in $10^3$.

Finally, we cross-checked results for the pure $N=1$ and $N=4$ cases using \textsc{Mathematica}, imposing boundary conditions that required (i) the slope of $X$ matches the slope of the Gaussian initial condition near the origin, and (ii) $X(\chi_\mathrm{max}, t)$ is equal to a large negative number (the results are insensitive to the exact value chosen).
Our results agree with those from the above method to within a percent.

\bibliography{Spectator}

\end{document}